\documentclass[aps,prd,preprint,nofootinbib,longbibliography]{revtex4-1}
\usepackage{blindtext}
\usepackage{mathtools}
\usepackage{xcolor}
\usepackage{amssymb,amsmath}
\usepackage{physics}
\usepackage{dsfont}
\usepackage{colortbl}

\usepackage[utf8]{inputenc}
\usepackage[english]{babel}
\usepackage{hyperref}
\begin{document}
\title{Space and Time Averaged Quantum Stress Tensor Fluctuations} \author{Peter Wu$^{1\dagger}$} \author{L. H. Ford$^{1\ddagger}$} \author{Enrico D. Schiappacasse$^{2,3*}$}
\affiliation{$^1$Institute of Cosmology, Department of Physics and Astronomy, Tufts University, Medford, Massachusetts 02155, USA\\
$^2$ Department of Physics, University of Jyv$\ddot{a}$skyl$\ddot{a}$, P.O. Box 35 (YFL), FIN-40014 Jyv$\ddot{a}$skyl$\ddot{a}$, Finland\\
$^3$Helsinki Institute of Physics, University of Helsinki, P.O. Box 64, FIN-00014 Helsinki, Finland}

\begin{abstract}
We extend previous work on the numerical diagonalization of quantum stress tensor operators in the Minkowski vacuum state, which considered operators averaged in a finite time interval, to operators averaged in a finite spacetime region.  Since real experiments occur over finite volumes and durations, physically meaningful fluctuations may be obtained from stress tensor operators averaged by compactly supported sampling functions in space and time.  
The direct diagonalization, via a Bogoliubov transformation, gives the eigenvalues and the probabilities of measuring those eigenvalues in the vacuum state, from which the underlying probability distribution can be constructed. 
For the normal-ordered square of the time derivative of a massless scalar field in a spherical cavity with finite degrees of freedom, analysis of the tails of these distributions confirms previous results based on the analytical treatment of the high moments.
We find that the probability of large vacuum fluctuations is reduced when spatial averaging is included, but the tail still decreases more slowly than exponentially as the magnitude of the measured eigenvalues increases, suggesting vacuum fluctuations may not always be subdominant to thermal fluctuations and opening up the possibility of experimental observation under the right conditions.
\end{abstract}

\maketitle

\section{Introduction}
The semiclassical theory of gravity, where matter fields are treated as quantum fields, whereas the gravitational field is treated as a classical field, deals with the expectation value of the energy-momentum tensor operator of the matter fields as an approximation to a full theory of quantum gravity~\cite{Ford:2005qz}.  This semiclassical theory offers a plausible description of the backreaction of the Hawking radiation on the gravitational background of a black hole~\cite{birrell_davies_1982} and 
opens the possibility of quantum particle creation~\cite{Schiappacasse:2016nei} through higher order derivative terms of the metric~\cite{PhysRevD.17.414}.

Nevertheless, the semiclassical theory does not provide a description of the expected quantum fluctuations of the stress tensor around its expectation value. Such fluctuations may have observable physical effects and, in recent years, has captured a great amount of attention from the physics community~\cite{Borgman:2003dm, Thompson:2006qe, Calzetta:1996sv, Ford:2010wd, Lombardo:2005iz, Wu:2006xp, Bessa:2014pya, Bessa:2016uqb}. Generally speaking, physical effects of the quantum fluctuations of a stress tensor operator may be addressed via the calculation of the probability distribution of the time- or spacetime-averaged operator.  Since these averages require normal ordering, the probability distribution in the vacuum state has zero mean and thus a nonzero probability of measuring negative components of the stress-energy tensor, such as the energy density.  In some sense, the averaging process may be viewed as a consequence of a physical measurement probing the outcomes of the operator.

The exact probability distribution associated with measurements of the stress-energy tensor in a two-dimensional conformal field theory in the vacuum is calculated in Ref.~\cite{Fewster:2010mc}. Using a Gaussian temporal sampling function, the resulting distribution is a shifted Gamma distribution, with the shift given by the optimal quantum inequality bound~\cite{Fewster:2004nj}. In four dimensions, the situation is more involved.  Qualitatively, the probability distributions may be inferred from the moments of the averaged operator, as done for several normal-ordered quadratic operators in the vacuum state using Lorentzian time averaging in Ref.~\cite{Fewster:2012ej} and later generalized in Ref.~\cite{Fewster:2015hga} for compactly supported functions, which are functions that are exactly zero outside a defined domain. The main prediction in both references is the asymptotic form of the distribution, which represents the probability of large fluctuations, falling more slowly than exponentially as
\begin{equation}
P(x) \sim c_0 x^b e^{-ax^c}\,,~x \gg1\,,
\label{eq:tasymptotic}   
\end{equation}
where $x$ refers to a dimensionless measurement of the stress tensor fluctuations and $a, b, c, c_0$ are constants that vary according to the smearing function and the specific operator. Take $u$ to be the Lorentzian time average of the electromagnetic energy density in a timescale $\tau$, so that a dimensionless measurement of the averaged energy density may be expressed as $x=u\tau^4$ in units where $\text{speed of light} = \hbar =1$. The asymptotic form for the probability of large fluctuations follow Eq.~(\ref{eq:tasymptotic}) with $a=1$ and $c=1/3$. Thus, large energy density fluctuations are more probable than one might have naively expected.  Smooth, compactly supported temporal sampling functions result in probability distributions that fall even more slowly~\cite{Fewster:2015hga}, because $c = \alpha/p$ in Eq.~(\ref{eq:tasymptotic}), where $0 < \alpha < 1$ and $p$ depends on the quantum operator. Here $\alpha$, to be defined in Sec. \ref{sec:probdisttheory}, is a parameter that describes the rate of switching. For example, for the energy density we have $p=3$. Note that compactly supported functions are more representative of measurements that occur in a finite period of time when compared to Lorentzian functions, which have an infinitely long tail.

An independent confirmation of the behavior of the tail of the probability distribution in Eq.~(\ref{eq:tasymptotic}) for the cases of Lorentzian and compactly supported temporal sampling functions is done in Ref.~\cite{Schiappacasse:2017oqu}. By performing a direct diagonalization of the time-averaged square of the time derivative of a massless scalar field in Minkowski spacetime, the authors numerically evaluate $P(x)$ for large vacuum fluctuations in a spherical cavity with finite degrees of freedom. The fitted values of the parameters $\{c_0, a, b, c\}$ that govern the asymptotic behavior of $P(x)$ are reported in Ref.~\cite{Schiappacasse:2017oqu} and are in good agreement with those obtained in Refs.~\cite{Fewster:2012ej, Fewster:2015hga} based on the moments approach.

The moments approach applied to quantum stress tensor operators averaged over finite time intervals is further developed in Ref.~\cite{PhysRevD.101.025006} to include averaging over finite regions of space.  If the spatial sampling scale is small compared to the temporal scale, the asymptotic behavior of $P(x)$ for a spacetime-averaged quadratic operator is expected to first decay as the worldline limit discussed earlier, with $c=\alpha/p$ in Eq.~(\ref{eq:tasymptotic}), before smoothly transitioning to a form with $c=\alpha$ instead.  Although the inclusion of spatial averaging increases the decay rate of the probability distribution compared to time averaging alone, the distribution still falls more slowly than exponentially for large $x$.  Under the right conditions, large vacuum fluctuations could then produce several physical effects that overshadow those of thermal fluctuations, opening up the possibility of experimental or observational confirmation.  In the following paragraphs, we briefly mention some of the most striking effects.

Fluctuating gravity waves produced by quantum stress tensor fluctuations of a conformal field in inflationary models are studied in Ref.~\cite{Wu:2011gk}. These gravity waves are potentially observable in the cosmic microwave background radiation and from gravity wave detectors, providing a probe of trans-Planckian physics.

Large vacuum radiation pressure fluctuations on particles with electric charge or nonzero polarizability may push the particles over potential barriers as shown in Ref.~\cite{Huang:2016kmx}. Depending on the details of the averaging over the finite time interval, the penetration rate via this mechanism may even surpass the known quantum tunneling rate.

The vacuum decay of a metastable state of a self-interacting scalar field is analyzed in Ref.~\cite{Huang:2020bzb}. Large quantum fluctuations of the time derivative of a scalar field averaged over a finite spacetime region lead to a decay rate comparable with the standard rate from the instanton approximation~\cite{PhysRevD.15.2929}. However, for operators that are quadratic in the time derivative of a scalar field, the probability distribution falls slower than an exponential function, in which case the decay rate is governed by these quadratic field fluctuations rather than quantum tunneling and linear field fluctuations.

Large fluctuations around the zero point density of a fluid as an analog model for quantum stress tensor fluctuations is studied in Ref.~\cite{Wu:2020hrz}. These density fluctuations may potentially be detectable in low-temperature light scattering experiments~\cite{Ford:2009zza} by observing fluctuations in the number of scattered photons.

In this paper, we extend the diagonalization approach developed in Refs. \cite{Schiappacasse:2017oqu, schiappacasse_2018} to treat probability distributions of quantum stress tensor operators averaged over a finite spacetime region.  The paper is organized as follows.  In Sec.~\ref{sec:probdisttheory}, we review the asymptotic behavior of the probability distributions of spacetime-averaged stress tensor operators, based on the high moments approach of Ref.~\cite{PhysRevD.101.025006}.  In Sec.~\ref{sec:diagonalization}, we consider the alternative diagonalization method, developed in Ref.~\cite{Schiappacasse:2017oqu}, that allows us to numerically construct the probability distributions.  We apply this numerical method to the square of the time derivative of a massless scalar field, and in Sec.~\ref{sec:setupuandresults} we discuss the approximations and analyze the results.  In Sec.~\ref{sec:conclusion}, we review the key takeaways and remaining loose ends.

Units in which the reduced Planck constant and the speed of light are equal to unity, $\hbar=c=1$, are used throughout the paper.

\section{Probability Distributions of Quantum Stress Tensor Operators}
\label{sec:probdisttheory}
The moments of a quantum operator can be used to infer the properties of the underlying probability distribution.  However, the moments of a quadratic field operator, which composes the stress tensor operator, are not well-defined at a single spacetime point, complicating efforts to do so.  One workaround is to investigate the moments of a quadratic field operator that has been averaged in time alone or space and time.\footnote{Note that quadratic operators averaged in space alone still have diverging moments in four spacetime dimensions, as discussed in footnote 2 of Ref.~\cite{Fewster:2015hga}.}  One is further led to consider averaging over a finite duration and volume, which are more representative of physical measurements in an experiment.  Consider a normal-ordered, quadratic field operator, which can be expanded in terms of creation and annihilation operators in the form
\begin{align}
\mathcal{T}(t,\mathbf{x}) = \frac{1}{2}\sum_{\mathbf{k},\mathbf{k'}} \Big[2a^\dagger_{\mathbf{k}}a_{\mathbf{k'}} F_{\mathbf{k}\mathbf{k'}}(t,\mathbf{x}) + a_{\mathbf{k}}a_{\mathbf{k'}} G_{\mathbf{k}\mathbf{k'}} (t,\mathbf{x}) +a^\dagger_{\mathbf{k}}a^\dagger_{\mathbf{k'}}G^*_{\mathbf{k}\mathbf{k'}} (t,\mathbf{x})\Big]\,.
\label{eq:tgeneralnoapprox}
\end{align}
The moments of the quadratic operator $\mathcal{T}(t,\mathbf{x})$ generically diverge.  In order to obtain finite moments, we need to average $\mathcal{T}(t,\mathbf{x})$ in time alone,
\begin{align}
    \overline{\mathcal{T}}(\mathbf{x})\equiv&\int_{-\infty}^\infty  dt \,f(t)\, \mathcal{T}(t,\mathbf{x})\,,\\
    =&\frac{1}{2}\sum_{\mathbf{k},\mathbf{k'}} \Big[2a^\dagger_{\mathbf{k}}a_{\mathbf{k'}} \overline{F}_{\mathbf{k}\mathbf{k'}}(\mathbf{x}) + a_{\mathbf{k}}a_{\mathbf{k'}} \overline{G}_{\mathbf{k}\mathbf{k'}} (\mathbf{x}) +a^\dagger_{\mathbf{k}}a^\dagger_{\mathbf{k'}}\overline{G}^*_{\mathbf{k}\mathbf{k'}} (\mathbf{x})\Big]
    \label{eq:tgeneralnoapproxtime}
\end{align}
to find
\begin{equation}
    \mu_n = \bra{0}\big[\,\overline{\mathcal{T}}(\mathbf{x})\,\big]^n\ket{0}\,,
    \label{eq:momentexpectationtime}
\end{equation}
or in space and time,
\begin{align}
    \overline{\mathcal{T}}\equiv&\int_{-\infty}^\infty dt \,f(t)
    \int_{\mathcal{V}} d^3x \,g(\mathbf{x})\,\mathcal{T}(t,\mathbf{x})\,,\\
    =&\frac{1}{2}\sum_{\mathbf{k},\mathbf{k'}} \Big[2a^\dagger_{\mathbf{k}}a_{\mathbf{k'}} \overline{F}_{\mathbf{k}\mathbf{k'}} + a_{\mathbf{k}}a_{\mathbf{k'}} \overline{G}_{\mathbf{k}\mathbf{k'}} +a^\dagger_{\mathbf{k}}a^\dagger_{\mathbf{k'}}\overline{G}^*_{\mathbf{k}\mathbf{k'}}\Big]
    \label{eq:tgeneralnoapproxspacetime}
\end{align}
to find
\begin{equation}
    \mu_n = \bra{0}\big[\,\overline{\mathcal{T}}\,\big]^n\ket{0}\,,
    \label{eq:momentexpectationspacetime}
\end{equation}
where $f(t)$ and $g(\mathbf{x})$ are the temporal and spatial sampling functions, respectively, and $\mu_n$ is the $n$th moment.  We assume the integrals of $f(t)$ in time and $g(\mathbf{x})$ in space are normalized to one, or equivalently that their Fourier transforms have the characteristic $\hat{f}(0)=\hat{g}(\mathbf{0})=1$.  The finite moments of the time-averaged or spacetime-averaged operators can be related to the moments of a probability density function,
\begin{align}
    \mu_n &= \int_{-\infty}^\infty dx \, x^n P(x)\,, \\
    &= \int_{-x_0}^\infty dx\,x^n P(x)\,, \label{eq:tprobmoment}
\end{align}
where $x$ denotes the eigenvalues of the operator in question, and the lower integral bound $-x_0$ comes from quantum inequalities.   Quantum inequalities are constraints on the expectation values of averaged stress tensor operators.  Because these expectation values can be arbitrarily negative when evaluated at a single spacetime point \cite{Epstein:1965zza}, macroscopic violations of physical laws become possible, a problem that can be resolved by arguing for quantum inequality constraints \cite{Ford:1978qya}.  Note that $-x_0$ is the lowest eigenvalue of the averaged operator, and hence is both the minimum expectation value and the lower bound on the probability distribution. 

The case for a time-averaged quadratic operator is discussed in detail in Ref.~\cite{Fewster:2015hga}, and here we proceed to summarize the main results.  Let us consider the normal-ordered quadratic operator
\begin{equation}
\mathcal{T}(t,\mathbf{x})=\tau^4(:\dot{\varphi}^2(t,\mathbf{x}):)\,.
\end{equation}
Here $\varphi(t,\mathbf{x})$ is the quantized massless scalar field, so $\dot{\varphi}^2(t,\mathbf{x})$ has dimensions of (length)$^{-4}$ in units where $\hbar=c=1$.  We introduce the extra factor of $\tau^4$ to make the operator dimensionless, where $\tau$ is the characteristic temporal sampling scale.

In rectangular coordinates, $\varphi(t,\mathbf{x})$ has the usual solution
\begin{equation}
    \varphi(t,\mathbf{x})=\sum_{\mathbf{k}}\frac{i}{2\omega V}\left(a_\mathbf{k}e^{i(\mathbf{k}\cdot\mathbf{x}-\omega t)}-a_\mathbf{k}^\dagger e^{-i(\mathbf{k}\cdot\mathbf{x}-\omega t)}\right)\,,
\end{equation}
where $V$ is the quantization volume and $\omega=k$.  Averaging $\mathcal{T}(t,\mathbf{x})$ in time, we find
\begin{equation}
    \overline{\mathcal{T}}(\mathbf{x})=\tau^4\int_{-\infty}^\infty dt \,f(t)
    \,(:\dot{\varphi}^2(t,\mathbf{x}):) \,.
    \label{eq:tdef}
\end{equation}
We assume $f(t)$ is a compactly supported, real, symmetric sampling function with a Fourier transform
\begin{equation}
    \hat{f}(\omega)=\int_{-\infty}^\infty dt\, f(t) e^{-i\omega t}
    \label{eq:ffourier}
\end{equation}
that asymptotically approaches
\begin{equation}
    \hat{f}(\omega)\sim C_f e^{-\beta|\omega\tau|^\alpha}\,, ~ |\omega\tau|\gg1\,.
    \label{eq:fhat}
\end{equation}
Here $C_f$ and $\beta > 0$ are constants and $0<\alpha<1$. Note that $\tau$, the characteristic sampling scale, may be defined by Eq.~(\ref{eq:fhat}) as the decay scale of the Fourier transform. For the functions which will be used in this paper, this is of the same order as the characteristic duration of $f(t)$, but this need not be true in general. Assuming a fixed spatial location ${\bf{x}={\bf{0}}}$, we find that $\overline{F}_{\text{{\bf{k}}},\text{{\bf{k}}'}}({\bf{x}={\bf{0}}})$ and $\overline{G}_{\text{{\bf{k}}},\text{{\bf{k}}'}}({\bf{x}={\bf{0}}})$ in Eq.~(\ref{eq:tgeneralnoapproxtime}) are given by  
\begin{align}
    \overline{F}_{\mathbf{k}\mathbf{k'}}(\mathbf{0}) &=\tau^4 \frac{\sqrt{\omega_{\mathbf{k}}\omega_{\mathbf{k'}}}}{V} \hat{f}(\omega_{\mathbf{k}}-\omega_{\mathbf{k'}})\,, 
    \label{eq:tavgf}\\
    \overline{G}_{\mathbf{k}\mathbf{k'}}(\mathbf{0}) &=\tau^4 \frac{\sqrt{\omega_{\mathbf{k}}\omega_{\mathbf{k'}}}}{V} \hat{f}(\omega_{\mathbf{k}}+\omega_{\mathbf{k'}}) \label{eq:tavgg} \,.
\end{align}
By calculating the moments in Eq. (\ref{eq:momentexpectationtime}), we find that, for large $n$, there is one dominant term given by the expression
\begin{equation}
    M_n=\sum_{\mathbf{k}_1 \dotsm \mathbf{k}_n} \overline{G}_{\mathbf{k}_1 \mathbf{k}_2} (\mathbf{0})\overline{F}_{\mathbf{k}_2 \mathbf{k}_3}(\mathbf{0}) \dotsm \overline{F}_{\mathbf{k}_{n-1} \mathbf{k}_n} (\mathbf{0})\overline{G}^*_{\mathbf{k}_n \mathbf{k}_1}(\mathbf{0})\,.
    \label{eq:momentdominant}
\end{equation}
Qualitatively, we may see this by noting that $\overline{F}_{\mathbf{k}\mathbf{k'}}(\mathbf{0})$ falls off more slowly than $\overline{G}_{\mathbf{k}\mathbf{k'}}(\mathbf{0})$ due to the arguments of $\hat{f}(\omega)$ in Eqs. (\ref{eq:tavgf}) and (\ref{eq:tavgg}). Since $a^\dagger_{\mathbf{k}}a_{\mathbf{k'}}\overline{F}_{\mathbf{k}\mathbf{k'}}(\mathbf{0})$ annihilates the vacuum state, we need $\overline{G}_{\mathbf{k}\mathbf{k'}}(\mathbf{0})$ and $\overline{G}^*_{\mathbf{k}\mathbf{k'}}(\mathbf{0})$ placed at the ends. $M_n$ contains the maximum possible number of factors of  $\overline{F}_{\mathbf{k}\mathbf{k}'}(\mathbf{0})$, which leads to its dominance over other terms in $\mu_n$. Taking the continuum limit, done in detail in Sec. IV of Ref.~\cite{Fewster:2015hga}, we find that $M_n$ when $n\gg 1$ is of the order
\begin{equation}
    M_n \sim \frac{3! C_f^2 [2\pi \tau f(0)]^{n-2}}{(2\pi^2)^n \alpha^5 (2\beta)^{(3n+2)/\alpha}} \Gamma\left[ \frac{(3n+2)}{\alpha}-4\right]\,.
    \label{eq:thighmoment}
\end{equation}
The Hamburger and Stieltjes moment theorems \cite{SIMON199882}, applied to distributions on whole lines and half-lines, respectively, guarantee unique probability distributions provided the moments do not grow too quickly with $n$.  For distributions that are bounded below, as in the case of stress tensor operators subject to quantum inequality constraints, the Stieltjes moment theorem may be more relevant.  The moments in Eq. (\ref{eq:thighmoment}) grow faster than the criteria of either moment theorem, so we cannot guarantee these moments specify a unique distribution.  However, it can be shown that distributions with moments given in Eq. (\ref{eq:thighmoment}) can be different from the previously referenced asymptotic form, Eq.~(\ref{eq:tasymptotic}), by merely some oscillatory function, leaving the salient features unaffected. Calculating the moments of the probability distribution in Eq.~(\ref{eq:tasymptotic}) using Eq.~(\ref{eq:tprobmoment}), we find
\begin{equation}
    \mu_n = \frac{c_0}{c} a^{-(n+b+1)/c} \Gamma\left[\frac{n+b+1}{c}\right]\,.
    \label{eq:tprobmomentspecific}
\end{equation}
Comparing Eqs. (\ref{eq:thighmoment}) and (\ref{eq:tprobmomentspecific}), we identify
\begin{equation}
    \begin{split}
    &a=2\beta\left[\frac{\tau f(0)}{\pi}\right]^{-\alpha/3},~ b=-\frac{4\alpha+1}{3}, ~ c=\alpha/3,\\
    & c_0 = ca^{(1+b)/c}3! C_f^2 \alpha^{-5}(2\beta)^{-2/\alpha}[2\pi \tau f(0)]^{-2}\,.
    \end{split}
    \label{eq:tparameter}
\end{equation}
Numerical simulations performed in Ref.~\cite{Schiappacasse:2017oqu} based on the direct diagonalization of the averaged operator $\overline{\mathcal{T}}(\mathbf{0})$ find good agreement with these predictions.

The moments approach may be readily extended for spacetime-averaged operators, which is discussed in Ref.~\cite{PhysRevD.101.025006}.  In this case we consider the spacetime-averaged analog of Eq. (\ref{eq:tdef}),
\begin{equation}
    \overline{\mathcal{T}}=\tau^4\int_{-\infty}^\infty dt \,f(t)
    \int_{\mathcal{V}} d^3x \,g(\mathbf{x})\,(:\dot{\varphi}^2(t,\mathbf{x}):) \,,
    \label{eq:tdefsa}
\end{equation}
where the choices of $\varphi(t,\mathbf{x})$ and $f(t)$ are identical to the time-averaged case, and $g(\mathbf{x})$ is a compactly-supported, real, spherically symmetric sampling function with a Fourier transform
\begin{equation}
    \hat{g}(\mathbf{k}) = \int_{\mathcal{V}} d^3x \, g(\mathbf{x}) e^{i\mathbf{k}\cdot\mathbf{x}}
\end{equation}
that asymptotically approaches
\begin{equation}
    \hat{g}(\mathbf{k})\sim \frac{C_g}{k^{2-\lambda}}e^{-\eta|\mathbf{k}\ell|^\lambda}\,, \; |\mathbf{k}\ell|\gg 1\,.
    \label{eq:ghat}
\end{equation}
Here $C_g$ is a constant, $\ell$ is the characteristic sampling length scale, and $0<\lambda<1$ with $\lambda \leq \alpha$.  Note that the factor of $k^{\lambda-2}$ arises in a specific function constructed in Ref. \cite{PhysRevD.101.025006}, but need not appear more generally.  The $\overline{F}_{\mathbf{k}\mathbf{k'}}$ and $\overline{G}_{\mathbf{k}\mathbf{k'}}$ matrix elements in Eq. (\ref{eq:tgeneralnoapproxspacetime}) are
\begin{align}
    \overline{F}_{\mathbf{k}\mathbf{k'}} &=\tau^4 \frac{\sqrt{\omega_{\mathbf{k}}\omega_{\mathbf{k'}}}}{V} \hat{f}(\omega_{\mathbf{k}}-\omega_{\mathbf{k'}}) \hat{g}(\mathbf{k}-\mathbf{k'})\,, \\
    \overline{G}_{\mathbf{k}\mathbf{k'}} &=\tau^4 \frac{\sqrt{\omega_{\mathbf{k}}\omega_{\mathbf{k'}}}}{V} \hat{f}(\omega_{\mathbf{k}}+\omega_{\mathbf{k'}}) \hat{g}(\mathbf{k}+\mathbf{k'}) \,.
\end{align}
The dominant contribution to the moments in Eq. (\ref{eq:momentexpectationspacetime}) is assumed to be the form given in Eq. (\ref{eq:momentdominant}) for the same reasons as the time-averaged case.  As in the time-averaged case, the high moments can be approximated and compared to the moments of the proposed $P(x)$ in Eq. (\ref{eq:tasymptotic}).   Interestingly, the Stieltjes moment theorem holds when $\alpha > 1/2$, suggesting a unique $P(x)$ can be determined in those cases.  The analytical calculation in Ref.~\cite{PhysRevD.101.025006}, though similar to that of the time-averaged case in Ref.~\cite{Fewster:2015hga}, is unable to precisely predict most of the parameters in Eq. (\ref{eq:tasymptotic}) due to poor understanding of the regime where the approximations hold.  The unambiguously predicted parameters in Eq. (\ref{eq:tasymptotic}) are $c$ and, with a caveat, $a$; the other parameters are not well-known.  The spacetime-averaged distribution is expected to eventually decay as
\begin{align}
    P(x)\sim \begin{cases}
    e^{-(x/B)^\alpha}\,, &~\lambda<\alpha\\
    e^{-[1+\eta(\ell/\tau)^\lambda](x/B)^\alpha}\,, &~\lambda=\alpha
    \end{cases}
    \text{ ~for }x \gg 1\,.
    \label{eq:spacetimeasymptotic}
\end{align}
Here $B$ is a constant predicted for a class of sampling functions in Sec. V D in Ref.~\cite{PhysRevD.101.025006} but otherwise not known in general.  Note that we are ignoring possible overall factors in powers of $x$ before the exponential in Eq. (\ref{eq:spacetimeasymptotic}), which are well predicted for the time-averaged case in Eq. (\ref{eq:tasymptotic}).  If $\ell < \tau$, the effect of spatial averaging is not pronounced for the lower moments, suggesting that the worldline behavior found in the time-averaged case approximately holds in some regime.  For a given $n$th moment of the probability distribution $P(x)$ in Eq.~(\ref{eq:tprobmoment}), one can estimate the location $x$ that contributes most to the integral, which depends on $n$.  Using an estimate of the highest moment for which the time averaging is dominant, we find that the worldline regime holds for $x \alt x_*$, where
\begin{equation}
x_*\sim \left(\frac{\tau \beta^{1/\alpha}}{\ell \eta^{1/\lambda}} \right)^3\,.
\label{eq:xstar}
\end{equation}
The asymptotic behavior of $P(x)$ for a spacetime-averaged quadratic operator is then expected to first decay as the worldline limit in Eqs. (\ref{eq:tasymptotic}) and (\ref{eq:tparameter}) until around $x\sim x_*$, where the distribution transitions to the form in Eq. (\ref{eq:spacetimeasymptotic}).

\section{Diagonalization of quadratic bosonic operators}
\label{sec:diagonalization}
Here we proceed to generalize the diagonalization procedure in Ref.~\cite{Schiappacasse:2017oqu} for time-averaged quantum stress tensor operators to include averaging over finite spatial volumes. 
\subsection{General procedure}
Our goal is to numerically evaluate $x$ and $P(x)$ for an arbitrary spacetime-averaged quadratic operator $\overline{\mathcal{T}}$ in the Minkowski vacuum state $\ket{0}_a$.  To do so, we need to solve for the eigenvalues of $\overline{\mathcal{T}}$ and corresponding probabilities of measuring those eigenvalues in the vacuum state, which amounts to a diagonalization problem.  

Recall that a generic spacetime-averaged operator can be expanded in the form given in Eq. (\ref{eq:tgeneralnoapproxspacetime}).  Here we will assume that $\overline{F}$ and $\overline{G}$ are real and symmetric matrices, so that
\begin{equation}
\overline{\mathcal{T}} = \frac{1}{2}\sum_{\mathbf{k},\mathbf{k'}} \Big[2a^\dagger_{\mathbf{k}}a_{\mathbf{k'}} \overline{F}_{\mathbf{k}\mathbf{k'}}
+\left(a_{\mathbf{k}}a_{\mathbf{k'}} +a^\dagger_{\mathbf{k}}a^\dagger_{\mathbf{k'}}\right)\overline{G}_{\mathbf{k}\mathbf{k'}}\Big]\,.
\label{eq:tgeneral}
\end{equation}
In general, the vacuum state will not be an eigenstate of $\overline{\mathcal{T}}$.  We perform a Bogoliubov transformation~\cite{Bogolyubov:1947zz} to convert $\overline{\mathcal{T}}$ into a diagonal form, with creation and annihilation operators $\{b_{\mathbf{k}},b_{\mathbf{k}}^\dagger\}$ acting on a different set of particle number states labeled by the subscript $b$, $\ket{\{n_{\mathbf{k}}\}}_b$, instead of $\{a_{\mathbf{k}},a_{\mathbf{k}}^\dagger\}$ acting on states $\ket{\{n_{\mathbf{k}}\}}_a$.  Such a transformation is done assuming $\{a_{\mathbf{k}},a_{\mathbf{k}}^\dagger\}$ can be written as a linear combination of $\{b_{\mathbf{k}},b_{\mathbf{k}}^\dagger\}$, which obey their own sets of commutation relations (see Sec.~III in Ref.~\cite{Schiappacasse:2017oqu} for further details).  Requiring the diagonal form
\begin{equation}
\overline{\mathcal{T}} = \sum_{\mathbf{k}} \lambda_{\mathbf{k}} b^\dagger_{\mathbf{k}}b_{\mathbf{k}} + C_{\text{shift}} \mathds{1}\,,
\end{equation}
where $\mathds{1}$ is the identity operator, the conditions outlined above give expressions for $\lambda_{\mathbf{k}}$ and $C_{\text{shift}}$, constants that depend on $\overline{F}_{\mathbf{k}\mathbf{k'}}$ and $\overline{G}_{\mathbf{k}\mathbf{k'}}$.  In analogy with ladder operators in quantum mechanics, we can express the $a$ vacuum state $\ket{0}_a$ in terms of the $b$ particle number states $\ket{\{n_{\mathbf{k}}\}}_b$ through clever use of the creation and annihilation operators.  Recall that the $a$ vacuum is the physical state in which we wish to study the fluctuations. It can be shown that
\begin{equation}
\ket{0}_a = \mathcal{N}e^{-\frac{1}{2}\mathbf{b}\mathcal{M}\mathbf{b}^{\dagger T}}\ket{0}_b\,.
\label{eq:avacuum}
\end{equation}
Here $T$ refers to the matrix transpose and $\mathbf{b}$ denotes a column matrix composed of $b_{\mathbf{k}}$'s.  The matrix $\mathcal{M}$ is derived by noting that $a_{\mathbf{k}}$, which can be written as a linear combination of $\{b_{\mathbf{k}},b_{\mathbf{k}}^\dagger\}$, annihilates the $a$ vacuum state $\ket{0}_a$, which itself is a linear combination of the $b$ number states $\ket{\{n_{\mathbf{k}}\}}_b$.  The constant $\mathcal{N}$ emerges from the usual normalization $_a\bra{0}\ket{0}_a=1$.  Doing the calculation in full, both $\mathcal{M}$ and $\mathcal{N}$ can be derived from $\overline{F}_{\mathbf{k}\mathbf{k'}}$ and $\overline{G}_{\mathbf{k}\mathbf{k'}}$.

Thus, for any vector in the eigenbasis $\ket{\{n_{\mathbf{k}}\}}_b$, the eigenvalue and corresponding measurement probability in the original $a$ vacuum state $\ket{0}_a$ is given by
\begin{align}
\overline{\mathcal{T}} \ket{\{n_{\mathbf{k}}\}}_b &= \sum_{\mathbf{k}} \left[\lambda_{\mathbf{k}} b^\dagger_{\mathbf{k}}b_{\mathbf{k}} + C_{shift} \mathds{1}\right]\ket{\{n_{\mathbf{k}}\}}_b\,,
\label{eq:eigenvalue}\\
P_{\{n_{\mathbf{k}} \}}& = \left| _b\bra{\{n_{\mathbf{k}}\}}\ket{0}_a\right|^2.
\label{eq:probability}
\end{align}
Equation (\ref{eq:probability}) can be numerically evaluated by expanding the exponential in Eq. (\ref{eq:avacuum}).

\subsection{Specific case: square of the time derivative of a massless scalar field}
We are interested in the specific case discussed earlier, with the spacetime-averaged operator $\overline{\mathcal{T}}$ given in Eq. (\ref{eq:tdefsa}) and the sampling functions behaving as discussed in Eqs. (\ref{eq:fhat}) and (\ref{eq:ghat}).  The Klein-Gordon equation for a massless scalar field is
\begin{equation}
    \Box \varphi(t,\mathbf{r})=0\,.
\end{equation}
Although Sec.~\ref{sec:probdisttheory} is done in rectangular coordinates, here we work in spherical coordinates to take advantage of the spherical symmetry of the spatial sampling function $g(r)$.  The solution for the positive frequency mode function is given by
\begin{equation}
    f_{\omega l m}(t,r,\theta,\phi)= \xi_{\omega l m} A_{\omega l m}e^{-i\omega t}P_l^m(\cos\theta)e^{i m \phi}j_l(kr)\,,
\end{equation}
where $\omega=k$, $j_l(r)$ are the spherical Bessel functions, $P_l^m(x)$ are the associated Legendre functions, $\xi_{\omega l m}$ is some phase factor, and $A_{\omega l m}$ is some constant to be determined.  A convenient choice for the phase factor is
\begin{equation}
    \xi_{\omega l m}=e^{i\pi(l+\frac{|m|}{2})}\,.
\end{equation}
Applying vanishing boundary conditions at the surface of a sphere of radius R, 
\begin{equation}
    f_{\omega l m}(t,r,\theta,\phi)\Bigr|_{r=R}=0\,,
\end{equation}
we find
\begin{equation}
    k_{nl}=\omega_{nl}=\frac{z_{nl}}{R}\,,
    \label{eq:omeganl}
\end{equation}
where $z_{nl}$ is the $n$th zero of the spherical Bessel function $j_l(kr)$.  Since the frequencies $\omega$ depend on $n$ and $l$, it is more convenient to label the solutions with $\{n,l,m\}$ instead of $\{\omega,l,m\}$.  Requiring that the commutation relations hold in second quantization, we find
\begin{equation}
    A_{nlm}=\sqrt{\frac{2l+1}{4\pi}\frac{(l-m)!}{(l+m)!}} \left(\sqrt{\omega_{nl}R^3}j_{l+1}(\omega_{nl}R)\right)^{-1}\,.
\end{equation}
For the Condon-Shortley phase convention, there is an extra factor of $(-1)^m$.  Expanding $\varphi(t,\mathbf{r})$ in terms of creation and annihilation operators,
\begin{equation}
    \varphi(t,\mathbf{r})=\sum_{n=1}^\infty \sum_{l=0}^\infty \sum_{m=-l}^l \Big[a_{nlm} f_{nlm}(t,\mathbf{r}) + a_{nlm}^\dagger f_{nlm}^*(t,\mathbf{r})\Big] \,.
\end{equation}
Differentiating in time, squaring the result, and ordering normally gives
\vspace{-0.1 cm}
\begin{widetext}
\begin{align}
   :\dot{\varphi}^2(t,\mathbf{r}): \, =\sum_{nlm}\sum_{n'l'm'} &\frac{\sqrt{\omega_{nl}\omega_{n'l'}}}{R^3} \frac{j_l(\omega_{nl}r) j_{l'}(\omega_{n'l'}r)}{j_{l+1}(\omega_{nl}R) j_{l'+1}(\omega_{n'l'}R)} \times\hspace{8 cm}\nonumber \\
    &\big(a^\dagger_{n'l'm'}a_{nlm} \xi_{lm}\xi^*_{l'm'} Y_{lm}(\theta,\phi)Y^*_{l'm'}(\theta,\phi) e^{-i(\omega_{nl}-\omega_{n'l'})t}\nonumber\\
    &\hspace{0.6 cm}-a_{nlm}a_{n'l'm'}\xi_{lm}\xi_{l'm'} Y_{lm}(\theta,\phi) Y_{l'm'}(\theta,\phi)e^{-i(\omega_{nl}+\omega_{n'l'})t} + \text{H.c.}\big)\,,
\label{eq:phidotsquare}
\end{align}
\end{widetext}
where H.c. refers to the Hermitian conjugate.  The spacetime average of $:\dot{\varphi}^2(t,\mathbf{r}):$ can be done by recalling the definition of the Fourier transform, Eq. (\ref{eq:ffourier}), and making use of the orthonormality conditions of the spherical harmonics.  We find, identically to Eq. (\ref{eq:tgeneral}), 
\begin{alignat}{2}
     \overline{\mathcal{T}} =\frac{1}{2}\sum_{\substack{nlm\\n'l'm'}} \big[2\overline{F}_{nlm,n'l'm'} a^\dagger_{nlm}a_{n'l'm'}
    + \overline{G}_{nlm,n'l'm'} (a^\dagger_{nlm}a^\dagger_{n'l'm'}+a_{nlm}a_{n'l'm'})\big]\,,
    \label{eq:tfinal}
\end{alignat}
where
\begin{equation}
\begin{split}
\overline{F}_{nlm,n'l'm'}=\frac{2\tau^4\sqrt{\omega_{nl}\omega_{n'l}}\,\delta_{l,l'}\, \delta_{m,m'}}{R^3j_{l+1}(\omega_{n'l}R)j_{l+1}(\omega_{nl}R)}\,\hat{f}\left(|\omega_{nl}-\omega_{n'l}|\right)
\int_0^{r_0} dr\, r^2 g(r) j_l(\omega_{nl}r)j_l(\omega_{n'l}r)
\label{eq:fmatrix}
\end{split}
\end{equation}
and
\begin{equation}
\begin{split}
\overline{G}_{nlm,n'l'm'}=\frac{-2\tau^4\sqrt{\omega_{nl}\omega_{n'l}}\,\delta_{l,l'}\, \delta_{m,-m'}}{R^3j_{l+1}(\omega_{n'l}R)j_{l+1}(\omega_{nl}R)}\,\hat{f}\left(\omega_{nl}+\omega_{n'l}\right)
\int_0^{r_0}dr\, r^2 g(r) j_l(\omega_{nl}r)j_l(\omega_{n'l}r)\,.
\label{eq:gmatrix}
\end{split}
\end{equation}
Here we have assumed that $g(r)$ has a compact support of $[0,r_0]$.  With the $\overline{F}$ and $\overline{G}$ matrices known, the procedure in Sec.~\ref{sec:diagonalization} can be performed to construct $P(x)$, which is done in Sec. \ref{sec:setupuandresults} for the case $n=1\sim600$, $l=m=0$.  In the limit of no spatial averaging, we expect to recover the purely time-averaged result.  Indeed, if we let
\begin{equation}
    g(\mathbf{r})=\delta^3(\mathbf{r})\,,
\end{equation}
we find
\begin{equation}
    \begin{split}
    \int d^3r \, \delta^3(\mathbf{r}) j_l(\omega_{nl} r) j_{l'} (\omega_{n'l'}r) &Y_{lm}(\theta,\phi)Y^*_{l'm'}(\theta,\phi)
    =\begin{cases}
    \frac{1}{4\pi}\,,~ &l=0\\
    0\,, &l\neq0
    \end{cases}
        \,.
    \end{split}
\end{equation}
The integral vanishes for $l\neq 0$ because $j_l(0)=0$ in those cases.  For $l=0$ we have $Y_{00}(\theta,\phi)=Y_{00}^*(\theta,\phi)$, so the result holds for all four terms in Eq. (\ref{eq:phidotsquare}).  We then get
\begin{equation}
\overline{F}_{n00,n'00}=\frac{(nn')^{3/2}\tau^4\pi^2}{2(-1)^{n+n'}R^4}\,\hat{f}\left(\frac{|n-n'|\pi}{R}\right)
\label{eq:fdelta}
\end{equation}
and
\begin{equation}
\overline{G}_{n00,n'00}=\frac{-(nn')^{3/2}\tau^4\pi^2}{2(-1)^{n+n'}R^4}\,\hat{f}\left(\frac{(n+n')\pi}{R}\right)\,,
\label{eq:gdelta}
\end{equation}
where it is understood that the matrix elements with $l\neq 0$ are zero.  Equations (\ref{eq:fdelta}) and (\ref{eq:gdelta}) are identical to the results in Ref.~\cite{Schiappacasse:2017oqu}, except for the extra factor of $1/(-1)^{n+n'}$, which arises due to a different choice of the phase factor $\xi_{nlm}$.

As discussed in Sec.~\ref{sec:probdisttheory}, the asymptotic behavior of $P(x)$ is primarily determined by the Fourier transforms of the sampling functions, so it can be useful to rewrite Eqs.~(\ref{eq:fmatrix}) and (\ref{eq:gmatrix}) in terms of $\hat{g}(k)$ instead of $g(r)$.  This can be done by calculating the Fourier transform using the plane wave expansion, giving
\begin{align}
    \int_0^{r_0}& dr\,r^2 g(r)  j_l(\omega r) j_l(\omega' r)
    = \frac{1}{8\pi}\int_{-1}^1 dx\, \hat{g}\left(\sqrt{\omega^2 + \omega'^2 -2\omega\omega' x}\right) P_l(x) \,,
    \label{eq:usehat}
\end{align}
where $P_l(x)$ are the Legendre polynomials.  Equation~(\ref{eq:usehat}) can be substituted into Eqs.~(\ref{eq:fmatrix}) and (\ref{eq:gmatrix}) so that the $\overline{F}$ and $\overline{G}$ matrix elements are written solely with the Fourier transforms of the sampling functions.

\section{Probability distribution function for a massless scalar field}
\label{sec:setupuandresults}
\subsection{Numerical setup}
\subsubsection{Construction and approximation of $\hat{f}(\omega)$}
\label{sec:timesampling}
A compactly supported time sampling function with a Fourier transform that asymptotically approaches the limit in Eq. (\ref{eq:fhat}) can be constructed following Sec.~IIB of Ref.~\cite{Fewster:2015hga}.  Let $\phi(t)$ be the inverse Laplace transform of $\tilde{\phi}(p)=e^{-(p\tau)^\alpha}$, where $0<\alpha<1$.  Defining $\hat{H}(\omega)$ to be the Fourier transform of $H(t)=\phi(t+\delta)\phi(t-\delta)$, the desired $\hat{f}(\omega)$ can be computed from 
\begin{equation}
\hat{f}(\omega)=\frac{\hat{H}^2(\omega)+\frac{1}{2}\left[\hat{H}^2(\omega+\frac{\pi}{2\delta})+\hat{H}^2(\omega-\frac{\pi}{2\delta})\right]}{\hat{H}^2(0)+\hat{H}^2(\frac{\pi}{2\delta})}\,,
\label{eq:fhatconstruction}
\end{equation}
where it can be shown that
\begin{align}
        C_f &= \frac{4\phi^2(2\delta)}{\hat{H}^2(0)+\hat{H}^2(\frac{\pi}{2\delta})}\,,\\
    \beta&=2\cos\left(\frac{\pi \alpha}{2}\right)\,.
\end{align}
Under this construction, the sampling function $f(t)$ has a compact support of $[-2\delta,2\delta]$.  The specification of the parameters $\{\alpha, \delta, \tau\}$ thus generates a particular time sampling function and corresponding Fourier transform.

Although we could perform the above procedure for the full set of $\{\omega\}$ in the computation, in practice there are a number of complications that motivate an approximation.  For a computation with many frequency modes, calculating $\hat{f}(\omega)$ point-by-point is time consuming and susceptible to numerical error at large $\omega$.  We also need to differentiate and integrate $\hat{f}(\omega)$, a task made easier with an analytic form.  We choose to approximate $\hat{f}(\omega)$ in the following way:
\begin{align}
  \hat{f}(\omega)=\begin{cases}
    \text{5th order spline interpolation}\,, & \omega \leq \omega_c\\
    C_f e^{-\beta|\omega\tau|^\alpha}\,, & \omega >\omega_c
  \end{cases}
  \,.
  \label{eq:hatfapprox}
\end{align}
The spline interpolation is performed on a sample dataset with $\omega\in[0,\omega_c]$.  When $\omega >\omega_c$, we directly evaluate the theoretically expected form, Eq. (\ref{eq:fhat}).  Here $\omega_c$ is chosen to be a point where the numerically computed $\hat{f}(\omega)$ approaches the theoretically expected limit but before any severe numerical error sets in.

\subsubsection{Construction of $g(r)$ or $\hat{g}(k)$}
\label{sec:spatialsampling}
A compactly supported, spherically symmetric spatial sampling function $g_1(r)$ that has a Fourier transform asymptotically approaching $\hat{g}_1(k)\sim e^{-(k\ell)^\lambda}$ can be found using the method in Sec. IID of Ref. \cite{Fewster:2015hga}.  Although this method was originally used to construct a one-dimensional temporal sampling function in Ref. \cite{Fewster:2015hga}, the argument holds for constructions of spherically symmetric spatial sampling functions, which take only a single argument $r$.  The asymptotic behavior of $\hat{g}_1(k)$ strongly depends on the properties of $g_1(r)$ near the end points, where $g_1(r)$ switches on and off.  Suppose we want $g_1(r)$ compactly supported in $r\in[0,r_0]$. Because we assume a spherically symmetric $g_1(r)$, the relevant behavior is the switch-on and switch-off as $r\to r_0^+$ and $r\to r_0^-$, respectively. Note that the function does not switch on or off at $r=0$, which is in the interior of the sampling region.

Crudely, to get $\hat{g}_1(k)\sim e^{-(k\ell)^\lambda}$, direct application of the method in Sec.~IID of Ref.~\cite{Fewster:2015hga} requires $g_1(r)$ to switch on as $e^{-r^{\lambda/(\lambda-1)}}$ as $r\to 0^+$.  In our case, since there is no switch-on at $r=0$, we instead want $e^{-(r_0-r)^{\lambda/(\lambda-1)}}$ as $r\to r_0^-$, which is just a reflection and translation to convert the switch-on at $r=0$ to a switch-off at $r=r_0$.
For the case $\lambda=0.5$, one option is then
\begin{align}
  g_1(r)=\begin{cases}
    0\,, & r \geq r_0\\
    A e^{-\frac{r_0}{r_0-r}}\,, & 0 \leq r < r_0
  \end{cases}
  \label{eq:gcoord}
\end{align}
Here $A=(3e)/\{2\pi r_0^3[8+13e\,\text{Ei}(-1)]\}$ is a normalization factor such that $\int d^3 r\,g(r) = 1$.  While this construction gives us a Fourier transform that asymptotically approaches $\hat{g}_1(k)\sim e^{-(k\ell)^\lambda}$, there are limitations to this method.  We do not have fine control over the Fourier transform itself, which is the most important function in the context of the high moments, as discussed in Sec.~\ref{sec:probdisttheory}.  In particular, the characteristic spatial sampling scale $\ell$ is unknown, complicating efforts to calculate the transition location $x_*$.  We are also not able to guarantee that the asymptotic limit goes exactly as Eq.~(\ref{eq:ghat}), which is the assumed behavior in Ref.~\cite{PhysRevD.101.025006}.  For the functions used in this paper, the characteristic decay length of $\hat{g}$ is of the same order as
the spatial sampling length.

For this reason, there are benefits to constructing the Fourier transform directly.  Here we consider the construction discussed in Ref.~\cite{PhysRevD.101.025006}.  Suppose we have a one-dimensional, compactly supported time sampling function $h(t)$ with a Fourier transform that asymptotically approaches
\begin{equation}
\hat{h}(\omega) \sim C_h e^{-\eta |\omega \tilde\tau|^\lambda}\,, \; \omega\tilde\tau \gg 1.
\label{eq:hhat}
\end{equation}
We now define a different spatial sampling function
\begin{equation}
    g_2(\mathbf{r}) = \frac{\tilde{\tau}^3h(|\mathbf{r}| \tilde{\tau}/\ell)}{2\pi \ell^3 |\hat{h}''(0)|}\,,
\end{equation}
for which the Fourier transform is
\begin{align}
\hat{g}_2(\mathbf{k}) &= \frac{\tilde\tau \hat{h}'(k\ell/\tilde\tau)}{k\ell\hat{h}''(0)}\,, \label{eq:ghatconstruct} \\[5pt]
&\sim \frac{\tilde\tau^2C_h\eta\lambda(kl)^{\lambda-2}}{|\hat{h}''(0)|}e^{-\eta(k\ell)^{\lambda}}\,, \; k\ell \gg 1.
\label{eq:ghatexpand}
\end{align}
Here $\hat{h}'(\omega)\equiv \frac{d}{d\omega} \hat{h}(\omega)$, and $\ell$ is now an input parameter we control.  Equation~(\ref{eq:ghatexpand}) can be found by explicitly taking the derivative of $\hat{h}(\omega)$ in Eq.~(\ref{eq:ghatconstruct}), using the asymptotic limit in Eq.~(\ref{eq:hhat}).  The construction given in Eq.~(\ref{eq:ghatconstruct}) has the same asymptotic behavior as Eq.~(\ref{eq:ghat}) once we identify
\begin{equation}
C_g= \frac{\tilde\tau^2C_h\eta\lambda \ell^{\lambda-2}}{{}|\hat{h}''(0)|}\,.
\end{equation}
Note that Eqs. (\ref{eq:fhat}) and (\ref{eq:hhat}) are identical: $\alpha$ and $\tau$ play the same roles as $\lambda$ and $\tilde\tau$, respectively.  The most straightforward choice for $\hat{h}(\omega)$ is then to choose $\hat{h}(\omega)=\hat{f}(\omega)$.  In this case, the compact support of $g_2(r)$ is $[0,2\delta\ell/\tau]$.

\subsubsection{Particle sectors and n,l,m}
\label{sec:nlm}
In principle, the diagonalization of a quadratic field operator calls for multiple infinite sums.  We may readily see this from Eq. (\ref{eq:tgeneral}), where a quadratic operator is expanded with creation and annihilation operators for all possible $\mathbf{k}$, and from Eqs. (\ref{eq:eigenvalue}) and (\ref{eq:probability}), where there are infinitely many $b$ particle number states $\ket{\{n_{\mathbf{k}}\}}_b$.  For this reason, we need to set upper bounds on these sums in such a way to preserve, as best possible, the fundamental structure of the probability distribution in the numerical implementation.  

Let us first consider Eq. (\ref{eq:tfinal}), which asks for infinite sums over $n,l,m$ and $n',l',m'$.  We would like the frequencies $\omega_{nl}$ to span a range as large as possible, to capture the contributions of small and large frequency modes.  Preliminary datasets have shown that the high frequency modes appear particularly important in generating data in the asymptotic region where Eq. (\ref{eq:spacetimeasymptotic}) is expected to hold.  The low frequency modes, on the other hand, appear to contribute larger probabilities, which are necessary for $P(x)$ to display the key decay features.  In light of Eq. (\ref{eq:omeganl}), we choose to fix $l=0$ and allow $n$ to span as wide a range as possible, i.e. to focus on the zeros of only the zeroth spherical Bessel function $j_0(kr)$.  The effect of other values of $l$ has not been investigated in depth, but here we work with $l=0$ because the zeros $z_{n0}$ grow the most slowly, which we anticipate will best capture contributions from both the low and high frequency modes given our computational constraints.  We thus consider a 600-mode setup with $n=1\ldots600$, $l=m=0$.  Although further increasing the range of $n$ would generate additional data at greater values of $x$, computations similar to those in this paper suggest the returns are marginal, especially on the log-log scales considered later.  A 600-mode setup gives us satisfactory data in a reasonable time frame, though there is nothing particular about this choice, and presumably different ranges of $n$ would work just as well.  Under this assumption, the square of the time derivative of the massless scalar field, Eq. (\ref{eq:tfinal}), becomes
\begin{equation}
    \overline{\mathcal{T}} =\frac{1}{2}\sum_{n,n'=1}^{600} \big[2\overline{F}_{nn'} a^\dagger_{n}a_{n'} + \overline{G}_{nn'} (a^\dagger_{n}a^\dagger_{n'}+a_{n}a_{n'})\big]\,,
    \label{eq:tfinall0}
\end{equation}
where, recalling that the $n$th zero of $j_0(kr)$ is $z_{n0}$=$n\pi$,
\begin{equation}
\begin{split}
\overline{F}_{nn'}=\frac{2\pi \tau^4 \sqrt{nn'}}{(-1)^{n+n'}R^2} \hat{f}\left(\frac{|n- n'|\pi}{R}\right)
\int_0^{r_0} dr\,g(r) \sin(\frac{n\pi r}{R})\sin(\frac{n'\pi r}{R})
\label{eq:fmatrixl0}
\end{split}
\end{equation}
and
\begin{equation}
\begin{split}
\overline{G}_{nn'}=\frac{-2\pi \tau^4 \sqrt{nn'}}{(-1)^{n+n'}R^2} \hat{f}\left(\frac{(n+n')\pi}{R}\right)
\int_0^{r_0} dr\,g(r) \sin(\frac{n\pi r}{R})\sin(\frac{n'\pi r}{R})\,.
\label{eq:gmatrixl0}
\end{split}
\end{equation}
The equation relating the integral with $g(r)$ to the integral with $\hat{g}(k)$, Eq. (\ref{eq:usehat}), can now be written in the simpler form
\begin{equation}
    \int_0^{r_0} dr\,g(r) \sin(\omega r)\sin(\omega' r) = \frac{1}{8\pi} \int_{|\omega-\omega'|}^{\omega+\omega'} dk\,k \hat{g}(k) \,.
    \label{eq:usehatl0}
\end{equation}
For the $l=0$ case, Eq. (\ref{eq:usehatl0}) can be more easily derived by doing the integral on the left-hand side in the complex plane.

The other upper bound to consider emerges from Eqs. (\ref{eq:eigenvalue}) and (\ref{eq:probability}), which ask for $b$ particle number states $\ket{\{n_{\mathbf{k}}\}}_b$ to calculate the eigenvalues and probabilities.  As there are infinitely many $b$ number states, we need to choose which states to include in the computation.  Let us first consider the $a$ vacuum state $\ket{0}_a$, which can be written in terms of $\ket{\{n_{\mathbf{k}}\}}_b$ using Eq. (\ref{eq:avacuum}) and a Taylor expansion of the exponential:
\begin{align}
\ket{0}_a &= \mathcal{N}\sum_{\rho=0}^\infty\left[\frac{\left( -\mathbf{b}^{\dagger} \mathcal{M} \mathbf{b}^{\dagger T}\right)^\rho}{\rho !}\right] \ket{0}_b\,, \\
&= \mathcal{N}\sum_{\rho=0}^\infty \left[\frac{1}{\rho !} \left(-\frac{1}{2}\sum_{i,j=1}^{600} b_i^{\dagger} \mathcal{M}_{ij}b_j^\dagger\right)^\rho\right]\ket{0}_b.
\label{eq:ketageneral}
\end{align}
The second sum in Eq. (\ref{eq:ketageneral}) is bounded above at $i,j=600$ due to our choice of $n=1\ldots600$, $l=m=0$.  The first sum over $\rho$ controls the number of $b$ and $b^\dagger$'s in the expansion of the $a$ vacuum state.  Note that the $b$ and $b^\dagger$'s always come in pairs, so the expansion of $\ket{0}_a$ in the basis of $b$ number states is such that only the even particle sectors of $\ket{\{n_{\mathbf{k}}\}}_b$ contribute: $\rho=0$ is the $b$ zero-particle sector, $\rho=1$ is the $b$ two-particle sector, and so on.  Because the particle number states are orthonormal, a choice of $\rho$ is merely a choice of states for which we are interested in computing the eigenvalues and probabilities.  Although an upper bound on the sum over $\rho$ is an approximation of the probability distribution $P(x)$, the eigenvalues and probabilities as calculated from Eqs. (\ref{eq:eigenvalue}) and (\ref{eq:probability}) are exact.  For example, suppose we want to calculate the probability of measuring, in the $a$ vacuum state, two particles in some $b$ number state, here labeled by $p$:
\begin{align}
    \left| _b\bra{2_p}\ket{0}_a\right|^2 &= \left|_b\bra{2_p} (-\frac{\mathcal{N}}{\sqrt{2}}M_{pp}\ket{2_p}_b + \cdots)\right|^2\,, \\
    &=\frac{\mathcal{N}^2}{2}\left|M_{pp} \right|^2\,.
\end{align}
In the expansion for $\ket{0}_a$, Eq. (\ref{eq:ketageneral}), the only term that matters is the one that is proportional to $\ket{2_p}_b$, because all other terms vanish when taking the inner product.

Now observe that the $2\rho$th particle sector contains a factor of $(\mathcal{M}_{ij})^\rho$ in Eq. (\ref{eq:ketageneral}).  The probabilities in the $2\rho$th sector will then go as $|\mathcal{M}_{ij}|^{2\rho}$.
Since $|\mathcal{M}_{ij}|<1$, for fixed $i$ and $j$ the higher particle sectors contribute smaller and smaller probabilities.  In light of the averaging process discussed later in section \ref{sec:averaging}, this suggests that the higher particle sectors can be ignored without affecting the probability distribution significantly.  In our setup we will calculate the eigenvalues and probabilities of only the 2-particle sector, corresponding to $\rho=1$.  However, note that in a typical computation $|\mathcal{M}_{ij}|$ for different $i$ and $j$ can span many orders of magnitude, so it is not true that the higher particle sectors always contribute negligible probabilities.  However, given the necessity for limiting the particle sectors in the computation, the most straightforward choice is to focus on the two-particle sector, which contributes significantly to the probability distribution across the board. Here it may be worth recalling the results in the worldline case. In Ref.~\cite{Schiappacasse:2017oqu}, Table I, some results for cumulative probability distributions are given. In all the cases studied there, the four-particle sector gives a contribution  of the order of $4\%$ or less of that of the two-particle sector. 

The two-particle sector also offers the advantage of being simpler to manage, as there are only two possible configurations of the momentum states.  If both particles are in the same momentum state, the eigenvalues and probabilities are given by
\begin{align}
    \overline{\mathcal{T}} \ket{2_i}_b &= (2\lambda_i + C_{\text{shift}})\ket{2_i}_b\,, \label{eq:2particleoutcome} \\
    P_{\{2_i\}} &=\frac{1}{2}|\mathcal{N}|^2|\mathcal{M}_{ii}|^2\,.
    \label{eq:2particleprob}
\end{align}
If the two particles are in different momentum states, we instead get
\begin{align}
    \overline{\mathcal{T}} \ket{1_i 1_j}_b &= (\lambda_i + \lambda_j+ C_{\text{shift}})\ket{1_i 1_j}_b\,,\\
    P_{\{1_i 1_j\}} &=|\mathcal{N}|^2|\mathcal{M}_{ij}|^2\,.
\end{align}
As we go to higher particle sectors, the number of configurations rises quickly, raising the additional question of which configurations to include in the computation, a problem we avoid with the two-particle sector.

\subsubsection{Averaging $P(x)$}
\label{sec:averaging}
The probability distribution $P(x)$ can be constructed by calculating the eigenvalues from Eq. (\ref{eq:eigenvalue}) and probabilities from Eq. (\ref{eq:probability}).  Note that these equations do not provide any \textit{a priori} reason to expect that $P(x)$ is a smooth distribution.  The arguments in Sec.~\ref{sec:probdisttheory} that lead to a smooth distribution rely on the moments of a quadratic operator, which do not necessarily encode the finer details of the distribution.  The analytical treatment assumes that the moments of a quadratic operator can be related to the moments of a smooth probability distribution function, which is a sensible conjecture but not proven.

Indeed, for a generic computation following Sec.~\ref{sec:diagonalization}, the constructed probability distribution is highly degenerate: for eigenvalues that are close together, the probabilities of measuring those eigenvalues can vary significantly.  The simplest way to see this is an extension of the argument in Sec.~\ref{sec:nlm}.  Higher $b$ number states are more likely to have smaller probabilities of being measured, but the outcomes of these measurements are not guaranteed to be much different from lower particle number states.  As an example, let us compare a measurement with two particles in the same momentum state with a measurement with four particles in the same momentum state.  The former case is given in Eqs. (\ref{eq:2particleoutcome}) and (\ref{eq:2particleprob}).  The analogous expressions for four particles are
\begin{align}
    \overline{\mathcal{T}} \ket{4_i}_b &= (4\lambda_i + C_{\text{shift}})\ket{4_i}_b\,,\\
    P_{\{4_i\}} &=\frac{3}{8}|\mathcal{N}|^2|\mathcal{M}_{ii}|^4\,.
\end{align}
One can imagine finding states $p$ and $p'$ for which the outcomes are similar,
\begin{equation}
    2\lambda_p + C_{\text{shift}} \simeq 4\lambda_{p'} + C_{\text{shift}}\,,
\end{equation}
but the probabilities are vastly different, 
\begin{equation}
    \frac{1}{2}|\mathcal{N}|^2|\mathcal{M}_{pp}|^2 \gg \frac{3}{8}|\mathcal{N}|^2|\mathcal{M}_{p'p'}|^4 \text{ (or vice versa).}
\end{equation}
In fact, to avoid the described situation across the infinitely many particle sectors would require much coincidence on the part of nature.  Perhaps such fine tuning in fact occurs in nature, but for a finite mode computation we need to deal with a degenerate $P(x)$.

In experimental and observational settings, we may not need to worry about these degeneracies. No realistic physical measurement probes a single eigenvalue $x$ of an operator, so the probability distribution $P(x)$ will always be integrated over some finite region of $x$.  We claim that the theoretical predictions in Sec.~\ref{sec:probdisttheory} are describing this physically observable probability distribution instead, one that has already been coarse-grained by the measurement process.  This coarse-grained distribution is the one we expect to be smooth and asymptotically approach the theoretical predictions.

Under this view, repeated measurements probing some range $\Delta x = x_f-x_i$ will find, on average,
\begin{equation}
    \bar{x} = \frac{\int_{x_i}^{x_f} dx\,x\,P(x)}{\int_{x_i}^{x_f} dx\, P(x)}\,,
\end{equation}
or, for the discrete probability distributions we are dealing with,
\begin{equation}
\bar{x} = \frac{\sum_{j} x_j\,P(x_j)}{\sum_{j} P(x_j)}\,.
\label{eq:xavg}
\end{equation}
Here the sum over $j$ is understood to be over all eigenvalues $x_j$ in the measurement domain, $x_j\in[x_i,x_f]$, and the denominator normalizes the total probability in this domain.  Note that Eq. (\ref{eq:xavg}) is merely the expectation value of $x$ in $[x_i,x_f]$, so we expect the probability of measuring $\bar{x}$ to be the mean of $P(x)$ over the same domain,
\begin{equation}
    \bar{P}(\bar{x}) = \frac{1}{\Delta x}\int_{x_i}^{x_f} dx\,P(x)\,,
    \label{eq:pavgprocedure}
\end{equation}
where the analogous expression for the discrete case is
\begin{eqnarray}
\bar{P}(\bar{x})=\frac{1}{\Delta x}\sum_{j} P(x_j)\,.
\label{eq:pavg}
\end{eqnarray}
Numerically, our prescription is to average the raw data using bins of width $\Delta x$, where the values of $\bar{x}$ and $\bar{P}(\bar{x})$ in each bin are given by Eqs. (\ref{eq:xavg}) and (\ref{eq:pavg}), respectively.\footnote{This binning procedure differs from Ref. \cite{Schiappacasse:2017oqu}.  The difference arises because here we work with the probability density function rather than the cumulative distribution function, which calls for separate considerations.  It can be shown, for example, that replacing the denominator in Eq. (\ref{eq:pavg}) with the number of points in the bin, as done in Ref. \cite{Schiappacasse:2017oqu}, does not guarantee the total probability is unity when integrating over all the bins.}  Doing so allows us to coarse-grain the raw data in a manner consistent with what we might expect from physical measurements.

The physical meaning of the bin sizes $\Delta x$ is not so clear.  One could argue that the sizes of these bins correspond to limiting factors in an experiment, such as the resolution of a detector or the uncertainty in the momentum of a photon probe, but such claims are purely speculative and would need to address the distinction between the averaging via binning discussed here and the spacetime averaging of the operators discussed in Section \ref{sec:probdisttheory}.  As the averaged data is fairly independent of the bin size $\Delta x$, our numerical simulations do not rely on particular choices of $\Delta x$.  In this sense we can also view the binning as a mathematical tool used to better analyze the numerical results and leave the trickier question of physical meaning for future investigation.

\subsection{Results}
Because the full asymptotic form of the probability distribution is not well-predicted, as shown in Eq. (\ref{eq:spacetimeasymptotic}) compared with Eq. (\ref{eq:tasymptotic}), our analysis needs to accommodate the undetermined parameters.  We would like to verify the unambiguous theoretical predictions: asymptotically, the averaged probability distribution will first decay as
\begin{equation}
    \bar{P}(\bar{x})\sim e^{-\bar{x}^{\alpha/3}}\,,
    \label{eq:worldlinelimit}
\end{equation}
before transitioning to a form that decays as
\begin{equation}
    \bar{P}(\bar{x})\sim e^{-\bar{x}^\alpha}\,,
        \label{eq:spacetimelimit}
\end{equation}
where the location of this transition is expected to occur at $x\sim x_*$, given in Eq. (\ref{eq:xstar}).  In Eqs. (\ref{eq:worldlinelimit}) and (\ref{eq:spacetimelimit}), it is understood that Eq. (\ref{eq:tasymptotic}) gives the full asymptotic form, of which we are concerned with the parameter $c$ that governs the exponential fractional decay rate.  Here we have written the equations using the averaged quantities $\bar{P}(\bar{x})$ and $\bar{x}$ to emphasize that we expect the theoretical predictions to hold for the coarse-grained probability distributions.  From Eq. (\ref{eq:tasymptotic}) we have
\begin{align}
\ln \bar{P}(\bar{x})&=\ln c_0 +b\ln \bar{x} -a\bar{x}^{c}\,, \label{eq:plog}\\
&\approx -a\bar{x}^{c}\,,
\label{eq:plotdropterms}
\end{align}
assuming $\bar{x}$ is sufficiently large.  Note that Eq. (\ref{eq:tasymptotic}) itself already assumes $\bar{x}\gg 1$, but here we require $\bar{x}$ to be even greater so that the third term in Eq. (\ref{eq:plog}) is dominant.  We thus find
\begin{equation}
\ln[-\ln \bar{P}(\bar{x})] \approx c\ln \bar{x} + \ln a.
\label{eq:plotapprox}
\end{equation}
A plot of $\ln[-\ln \bar{P}(\bar{x})]$ against $\ln \bar{x}$ then gives a slope of $c$, which is $\alpha/3$ in the worldline limit, Eq. (\ref{eq:worldlinelimit}), and $\alpha$ in the spacetime-averaged limit, Eq. (\ref{eq:spacetimelimit}).

\begin{figure}[ht]
\centering
\includegraphics[scale=0.58]{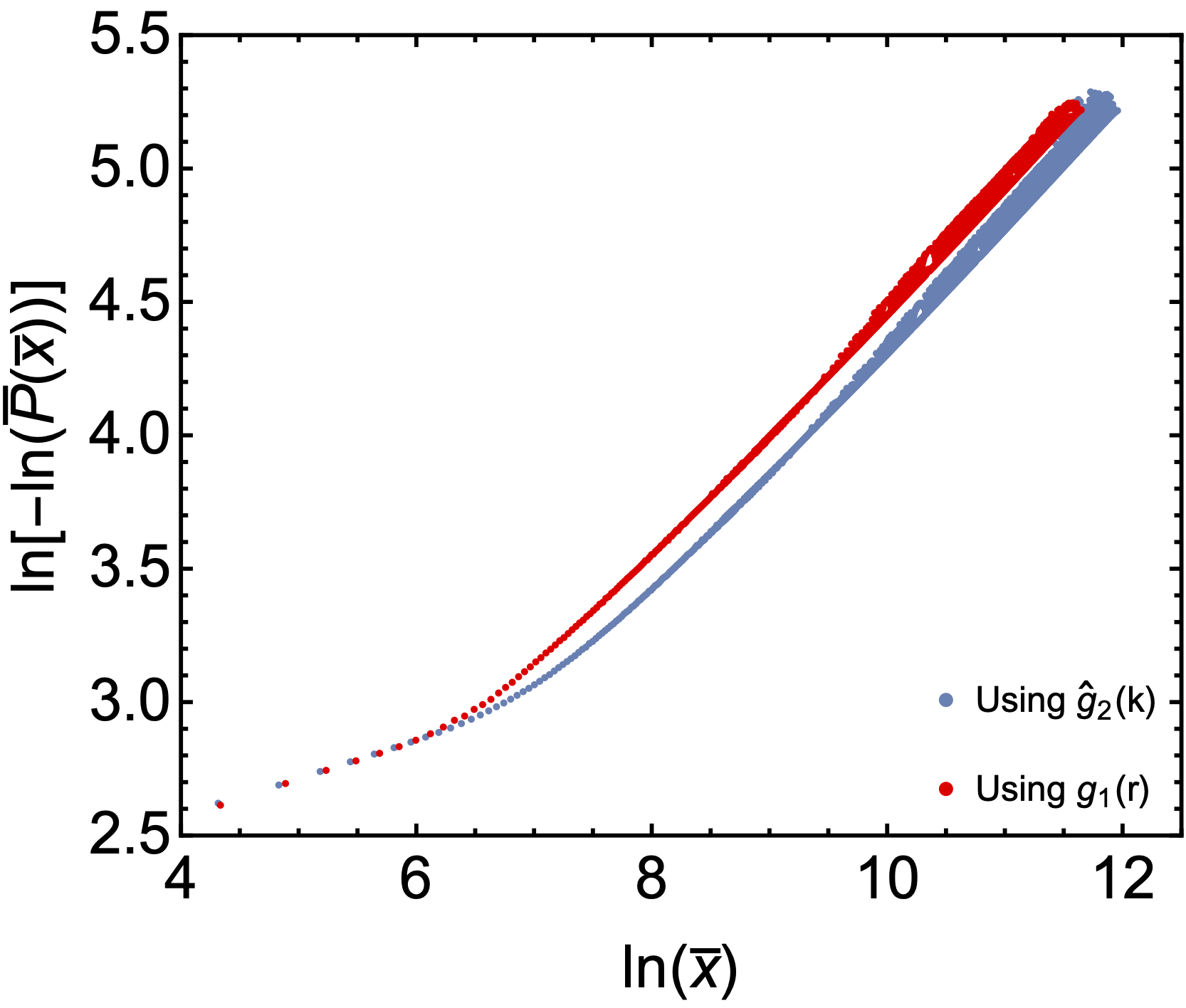}\!\!\!\!\!\!\!\!\!\!\!\!\!
\caption{A plot comparing the numerical results using $\hat{g}_2(k)$ and $g_1(r)$, defined in Eqs. (\ref{eq:ghatexpand}) and (\ref{eq:gcoord}), respectively, for the case $\alpha=\lambda=0.5$ and an averaging bin size $\Delta x =50$.  Here we have kept the compact supports of both functions in coordinate space, $g_1(r)$ and $g_2(r)$, the same, with $r_0=2\delta \ell /\tau = \ell = 0.28$, $\tau=1$,} and $\delta=0.5$.
\label{fig:comparison}
\end{figure}

As discussed in Sec. \ref{sec:spatialsampling}, it is more advantageous to work with $\hat{g}_2(k)$ instead of $g_1(r)$ to allow for finer control over the Fourier transform and better compatibility with Ref. \cite{PhysRevD.101.025006}.  However, in principle the numerical diagonalization can be performed using either function, as shown in Fig. \ref{fig:comparison}, where the averaged probability distributions are remarkably similar despite $g_1(r)\neq g_2(r)$.  In particular, we see that two sampling functions with similar asymptotic behavior in Fourier space do indeed produce similar probability distributions, lending credence to the claim that the Fourier transforms govern the tail region of the probability distributions.

We work with datasets that perform the numerical diagonalization using the Fourier transforms of the sampling functions, $\hat{f}(\omega)$ and $\hat{g}_2(k)$, and we consider the cases $\alpha=\lambda=0.5$; $\alpha=0.7$, $\lambda=0.5$; and $\alpha=\lambda=0.7$.  The parameters used in the construction of $\hat{f}(\omega)$ and $\hat{g}_2(k)$ are shown in Table \ref{tab:parameters}, where we explicitly work in $\tau=1$ units.  In principle, we would like to take the limit where the boundary of the sphere is infinite, i.e., $R\to\infty$.  Because we are limited to a finite number of modes, we are forced to consider a finite boundary.  A compromise is to take $2R$ or $4\pi R^3/3$ to be greater than the sampling times and volumes, respectively, so that the presence of the boundary would not be observable in a measurement.  We satisfy this requirement for our sampling volumes, but numerical instabilities do not allow us to choose small enough sampling times without sacrificing the high frequency modes.  The latter problem is also noted in Ref. \cite{Schiappacasse:2017oqu}, and as in that publication we acknowledge that the total sampling duration, $4\delta$, is greater than $2R$, the distance to travel from the origin to the boundary and back, or from one end of the boundary to the opposite end.  The simulations may thus be an imperfect approximation of Minkowski space, but we expect the setup to be approximate enough to compare with the theoretical calculations.

For each of the three cases, the averaging bin size $\Delta x$ is chosen to reduce scatter in the spacetime-averaged\hspace{0,1 cm} regime without\hspace{0,1 cm} smearing out\hspace{0,1 cm} the\hspace{0,1 cm} worldline behavior
completely.\hspace{0,1 cm} For 

\begin{table*}[hbt]
\begin{center}
\caption{Parameter choices used for the construction of $\hat{f}(\omega)$. As discussed in Sec. \ref{sec:timesampling}, a choice of $\{\alpha,\delta,\tau\}$ determines $\beta$ and $C_f$.  Recall from Sec. \ref{sec:spatialsampling} that $\hat{g}_2(k)$ is constructed using $\hat{h}(\omega)=\hat{f}(\omega)$, with $\lambda=\alpha$, $\eta=\beta$, $\tilde\tau=\tau$, and $C_h=C_f$.  The value of $|\hat{f}''(0)|$ is numerically evaluated from the spline interpolation in Eq. (\ref{eq:hatfapprox}).}
\vspace{0.2cm}
\begin{tabular}{rrrrrr}
\hline
$\alpha$ ~~~~~& $\beta$ ~~~~~~&$\tau$~~&$\delta$~~~&$C_f$ ~~~~~& $|\hat{f}''(0)|$\\          
\hline
0.5\textcolor{white}{999}& 1.41\textcolor{white}{999}&1\textcolor{white}{9}&0.5\textcolor{white}{9}& 2.9324\textcolor{white}{99} & 0.0763\\
0.7\textcolor{white}{999}& 0.908\textcolor{white}{99}&1\textcolor{white}{9}&1\textcolor{white}{9}&0.5235\textcolor{white}{99}& 0.253\textcolor{white}{9}\\
\hline
\label{tab:parameters}
\end{tabular}
\end{center}
\end{table*}
\vspace{-1.5 cm}
\begin{table*}[ht]
\begin{center}
\caption{Fit results for the three cases.  Numerical instabilities lead to the different choices of the spherical boundary $R$, but we keep the ratio $\ell/R$ constant. Here the acronyms WL and STA denote worldline and spacetime-averaged, respectively.}
\vspace{0.0cm}
\begin{tabular}{rrrrrrrrr}
\hline
$\text{Case}\textcolor{white}{9999}$&$R$\textcolor{white}{9}& $\ell\textcolor{white}{99}$&$\Delta x$&$\text{Predicted} \ln x_*$&$\text{Regime}$ &$ \text{Predicted slope}$ & $\text{Fitted slope}$ & $\text{Standard error}$\\
\hline
$\alpha=\lambda=0.5$~~~& 0.88&0.14&50&5.90\textcolor{white}{99999}&WL\textcolor{white}{9}& 0.167\textcolor{white}{9999}&0.1404\textcolor{white}{99}&0.0004\textcolor{white}{9999}\\
&&&&& STA& 0.5 \textcolor{white}{9999}&0.5016\textcolor{white}{99}&0.0012\textcolor{white}{9999}\\
\hline
$\alpha=0.7$, $\lambda=0.5$~~~& 1.57&0.25&10&1.67\textcolor{white}{99999}&WL\textcolor{white}{9}&0.233\textcolor{white}{9999}&0.1739\textcolor{white}{99}&0.0009\textcolor{white}{9999}\\
&&&&&STA& 0.7 \textcolor{white}{9999}&0.6925\textcolor{white}{99}&0.0013\textcolor{white}{9999}\\
\hline
$\alpha=\lambda=0.7$~~~& 1.57&0.25&0.5&4.16\textcolor{white}{99999}&WL\textcolor{white}{9}&0.233\textcolor{white}{9999}&0.1921\textcolor{white}{99}&0.0018\textcolor{white}{9999}\\
&&&&&STA& 0.7\textcolor{white}{9999}&0.7010\textcolor{white}{99}&0.0008\textcolor{white}{9999}\\
\hline
\label{tab:results}
\end{tabular}
\end{center}
\end{table*}

\hspace{-0.5 cm}this reason, datasets for which the worldline behavior ends earlier require smaller $\Delta x$.  As we do not expect the measurement outcomes $x$, and thus the averaged outcomes $\bar{x}$, to be correlated, a least-squares linear fit is sufficient for our purposes.  However, in order to account for the different number of data points in each bin, we perform weighted least-squares linear fits to find the slopes.  As a shorthand, let us write
\begin{equation}
    z_i \equiv \ln[-\ln \bar{P}(\bar{x}_i)]
    \label{eq:zi}
\end{equation}
and
\begin{equation}
    y_i \equiv \ln{\bar{x}_i}.
\end{equation}
We want a linear fit to Eq. (\ref{eq:plotapprox}), which can now be written as
\begin{equation}
    z = \gamma_1 y + \gamma_2,
\end{equation}
where the expected values for $\gamma_i$ are given in Eq. (\ref{eq:plotapprox}): $\gamma_1=d$, $\gamma_2=\ln a$.  We estimate $\gamma_i$ by finding the parameters $\tilde{\gamma}_i$ such that the weighted squared residuals
\begin{equation}
    s^2 = \sum_i w_i \left[z_i-\tilde{z}(y_i;\tilde{\gamma}_1,\tilde{\gamma}_2)\right]^2
    \label{eq:minimization}
\end{equation}
are minimized, where $\tilde{z}(y;\tilde{\gamma}_1,\tilde{\gamma}_2)$ is a linear fit function.  Here $w_i$ is the weight associated with the $i$th squared residual; for non-weighted least-squares fits the conventional choice is to take $w_i=1$ for all $i$.  In our numerical computation, $w_i$ is taken to be the number of points in the $i$th bin divided by the total number of points.

\begin{figure}[ht!]
    \centering
    \begin{minipage}[c]{\linewidth}
    \includegraphics[scale=0.54]{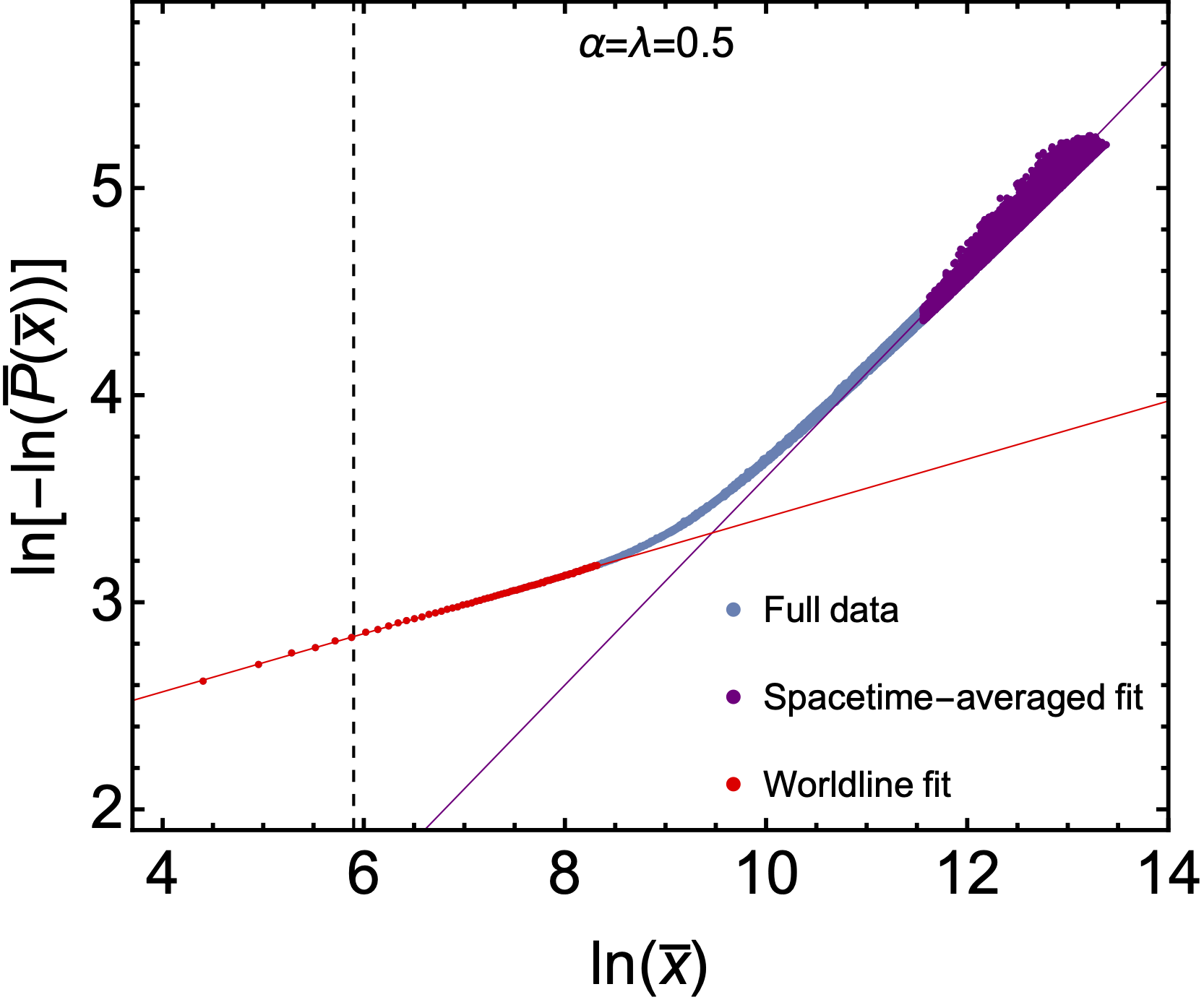}
    \end{minipage}
    \\
    \vspace{2mm}
    \begin{minipage}[c]{\linewidth}
    \includegraphics[scale=0.54]{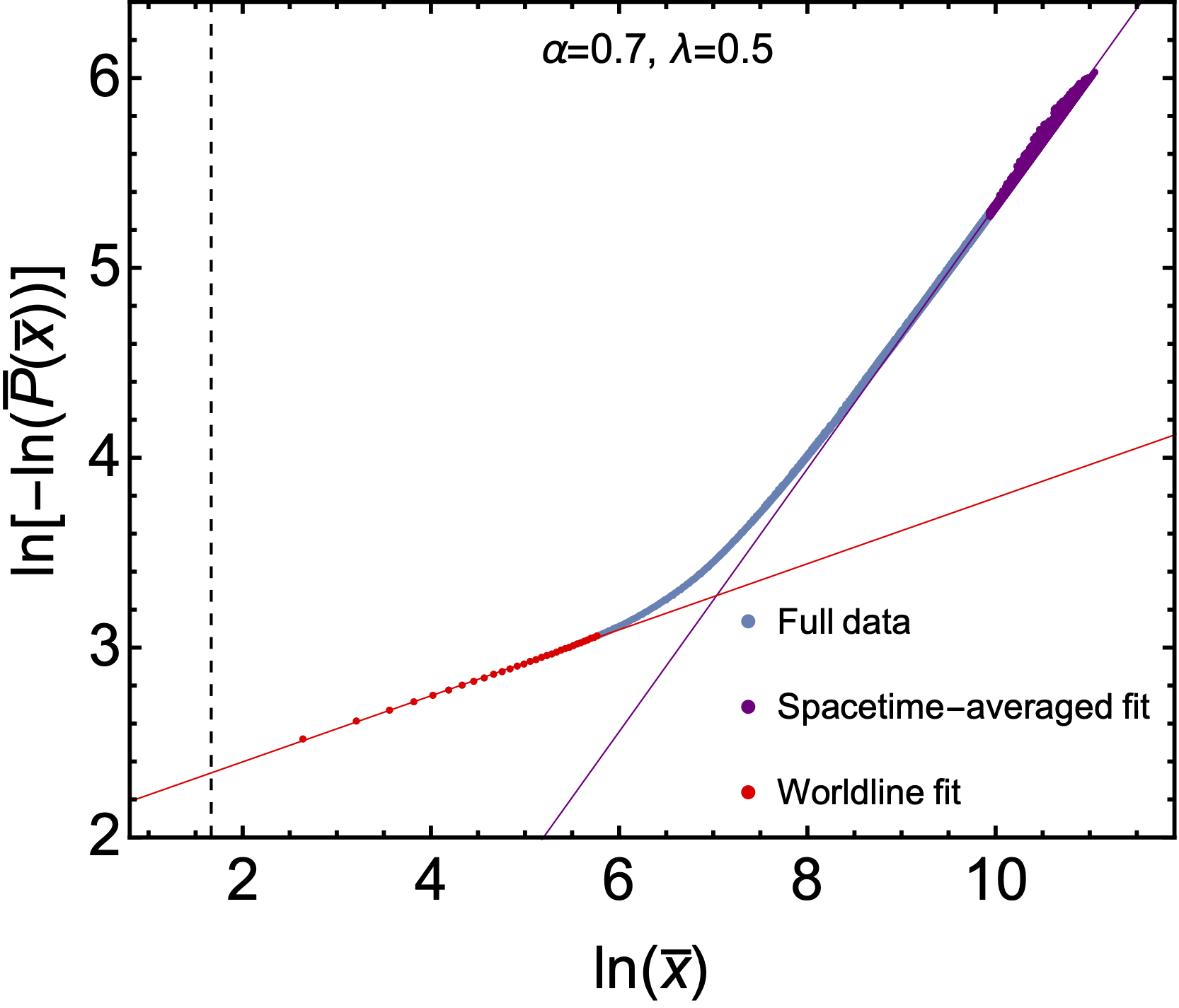}
    \end{minipage}
    \\
    \vspace{2mm}
    \begin{minipage}[c]{\linewidth}
    \includegraphics[scale=0.54]{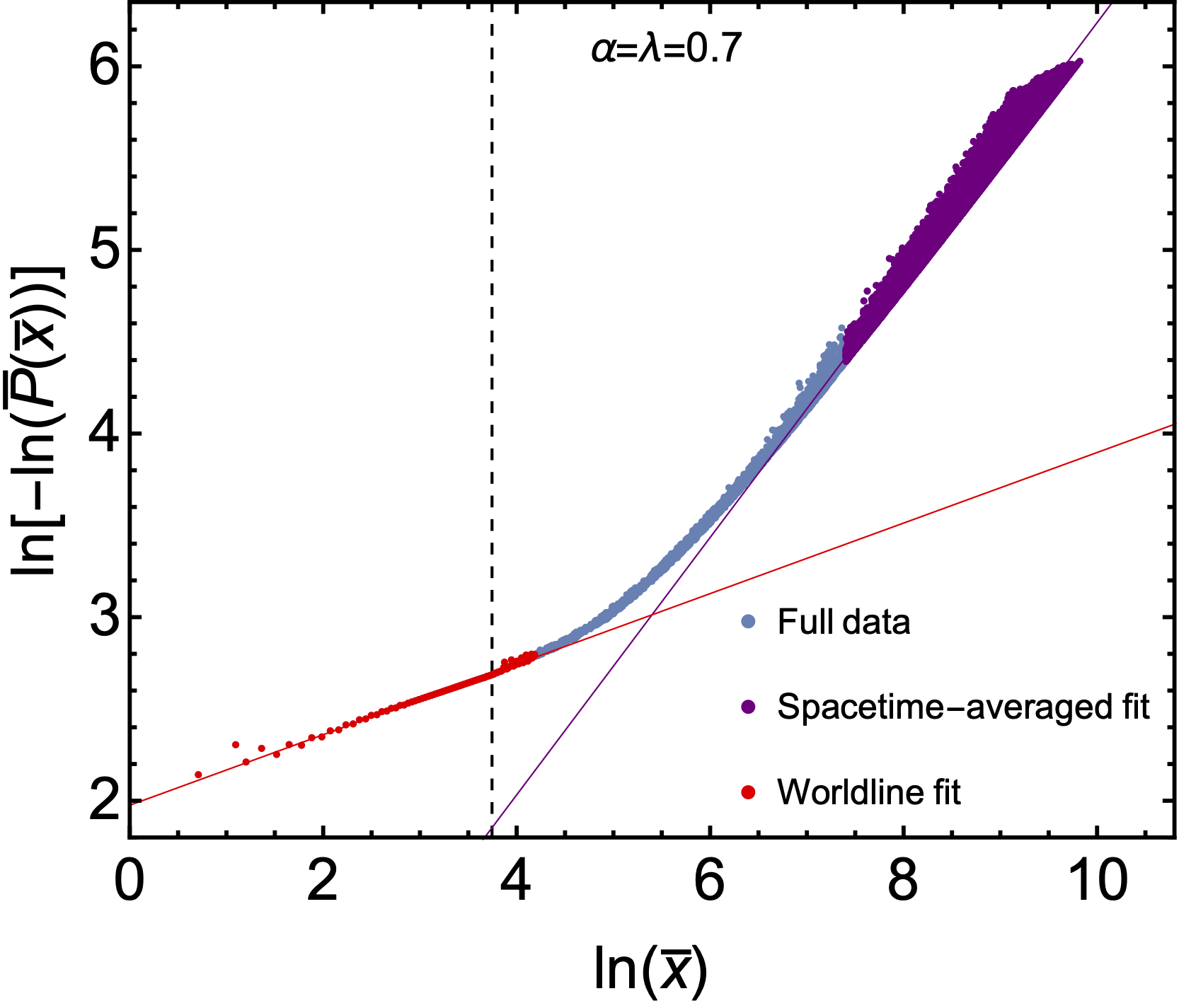}
        \caption{Best-fit lines in the worldline and spacetime-averaged regimes, where the bins used for the fits are colored accordingly.  The vertical dashed lines denote the predicted transitions between the two limits, $\ln x_*$, from Eq.~(\ref{eq:xstar}).  The parameters used for each dataset are given in Table \ref{tab:results}.}
            \label{fig:results}
    \end{minipage}
\end{figure}

\begin{figure}[ht!]
\centering
\includegraphics[scale=0.58]{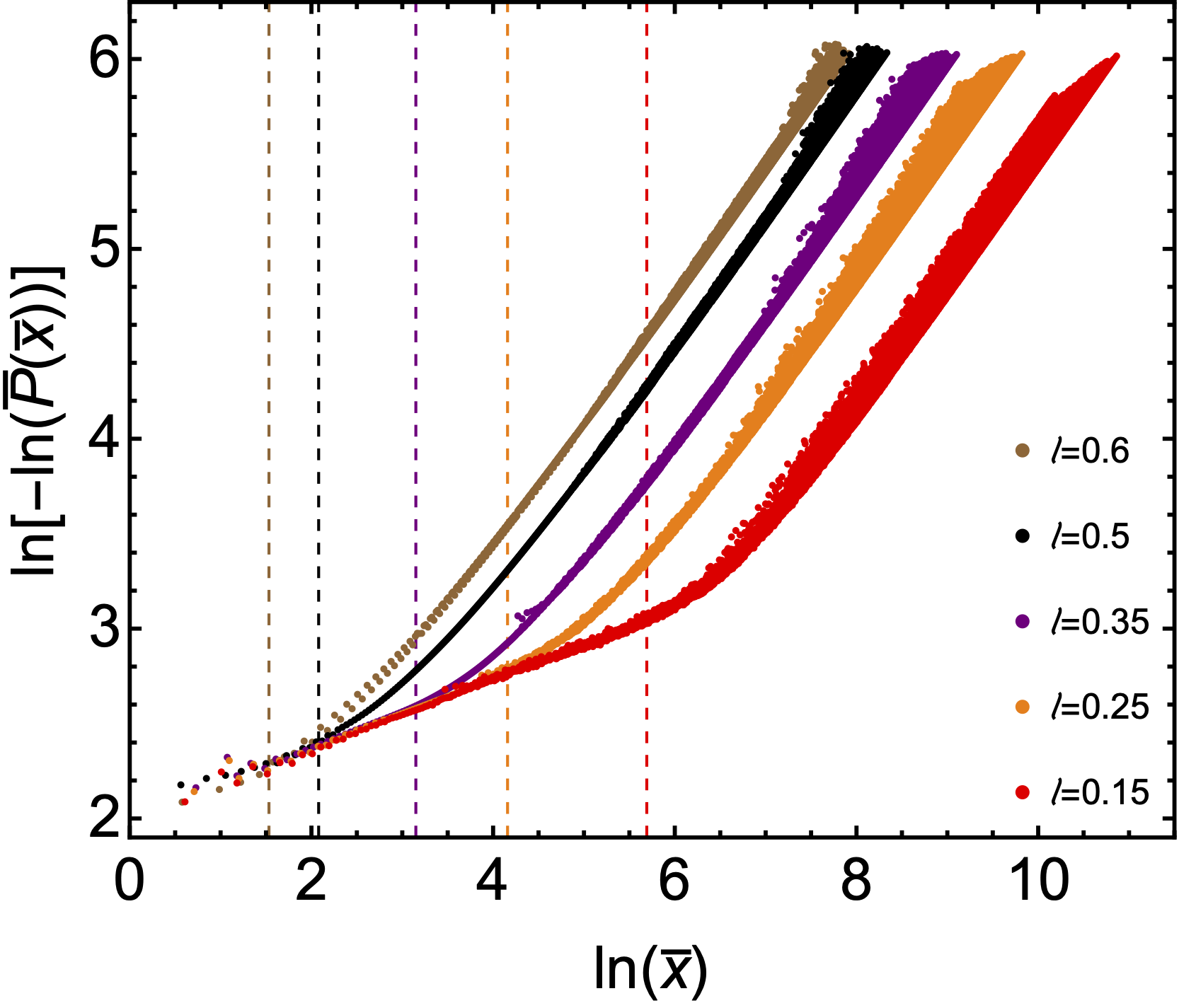}\!\!\!\!\!\!\!\!\!\!\!\!\!
\caption{Numerical results with $\alpha=\lambda=0.7$, where the characteristic spatial sampling length $\ell$ is varied.  In all datasets, the averaging bin size is $\Delta x =0.5$.  The colored dashed lines are the predicted transition locations, $\ln x_*$, for the correspondingly colored datasets.  These transitions are estimated from Eq.~(\ref{eq:logpredict}) with the parameters given in Table I.}
\label{fig:transition}
\end{figure}

The fit results are compiled in Table \ref{tab:results} and Fig.~\ref{fig:results}.  The numerical results for the exponential decay rates in the spacetime-averaged limits match the predictions exceedingly well.  In contrast, the decay rates in the worldline limits are lower than expected in all cases.  One explanation is the large $\bar{x}$ approximation in Eq. (\ref{eq:plotdropterms}) may not hold for our worldline data in the region $\ln \bar{x} = 4\sim 9$.  Moreover, as shown in Fig. \ref{fig:transition}, datasets with greater $\ell$ have less data in the worldline region, limiting the reliability of our fits.  Nonetheless, the worldline limit has been extensively investigated in Ref.~\cite{Schiappacasse:2017oqu}, where numerical simulations verified the full asymptotic form in Eqs. (\ref{eq:tasymptotic}) and (\ref{eq:tparameter}), so the poorer worldline results here are likely due to limitations in our analysis or data sets.

As shown Fig. \ref{fig:transition}, the predicted transition locations from worldline to spacetime-averaged behavior are underestimates, suggesting the worldline behavior is more robust than predicted.  We see that the transition behavior qualitatively holds, as datasets with smaller $\ell$ do transition at greater $\bar{x}$, though the calculations leading to Eq. (\ref{eq:xstar}) may be too rough to accurately pinpoint the locations of these transitions.

\begin{figure}[ht!]
\centering
\includegraphics[scale=0.58]{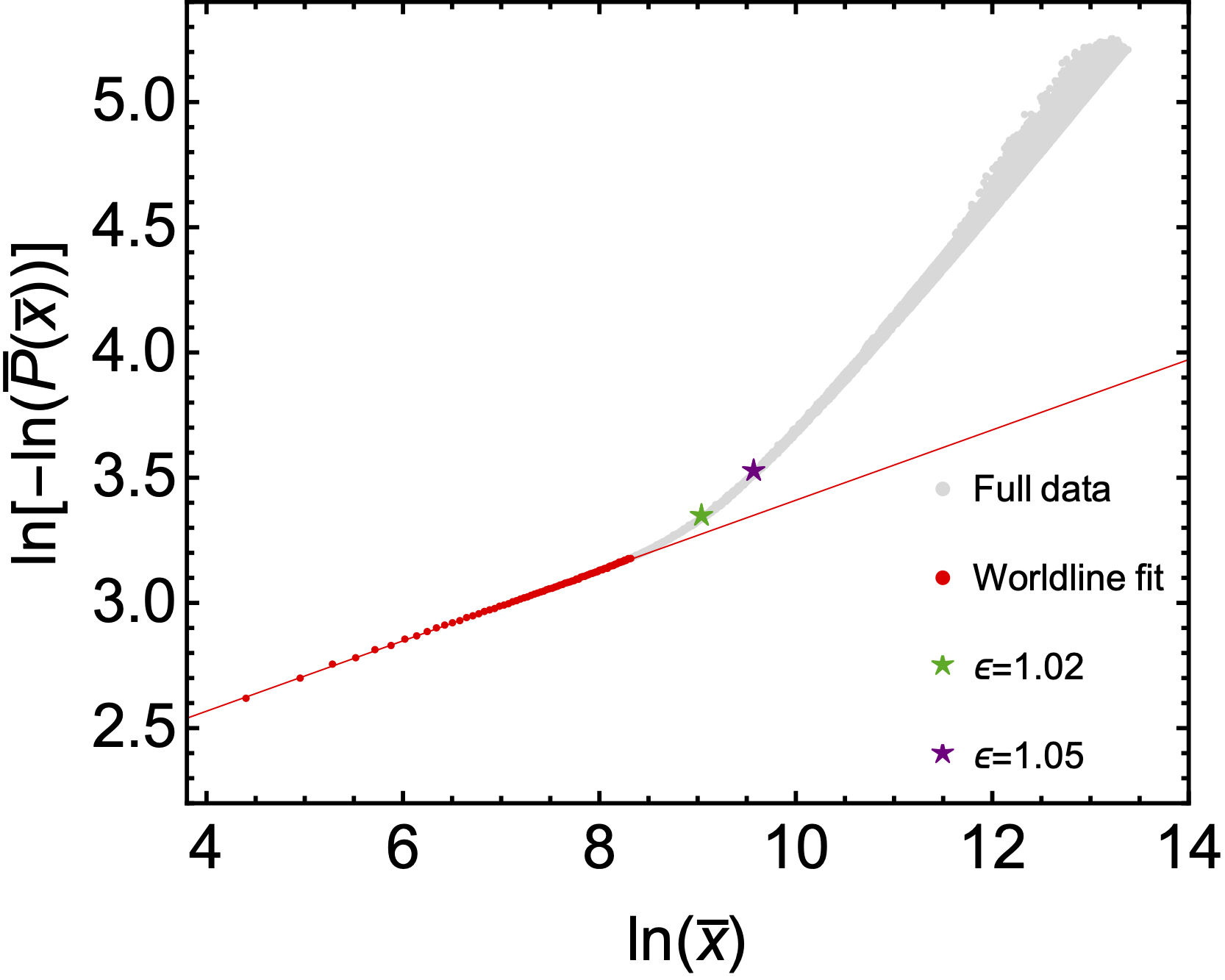}\!\!\!\!\!\!\!\!\!\!\!\!\!
\caption{Plotted are the numerical estimates of $\ln x_*$, depicted as green and purple stars, for two choices of the threshold $\epsilon$.  The dataset here uses $\alpha=\lambda=0.5$ and $\ell=0.14$, identical to that plotted in Fig. \ref{fig:results}.}
\label{fig:threshold}
\end{figure}

We can consider a more detailed analysis of the transition behavior by noting that
\begin{equation}
\ln{x_*} \sim 3\ln{\left(\frac{\tau \beta^{1/\alpha}}{\ell \eta^{1/\lambda}}\right)}\,,
\label{eq:logpredict}
\end{equation}
which can be numerically probed via linear fits.  The actual transition locations can be numerically estimated through a number of methods, one of which we discuss here.  Theoretically, $x_*$ is predicted to be the point at which the worldline behavior ends.  At this point, by definition the data  will deviate from the linear fits in the worldline region.  Following the notation in Eqs. (\ref{eq:zi})-(\ref{eq:minimization}), we can estimate these points by taking $\ln x_*$ to be the smallest $y_i$ where
\begin{equation}
\frac{z_i}{\tilde{z}(y_i)} \geq \epsilon\,,
\label{eq:threshold}
\end{equation}
provided $(y_i,z_i)$ is not a data point used in the linear fit, which has been assumed to be part of the worldline regime.  Here $\epsilon$ is a constant that sets the threshold for how much the data need to deviate from the worldline behavior, so $\epsilon>1$.  Some estimates of $\ln x_*$ are given in Fig. \ref{fig:threshold}, where we observe that larger values of $\epsilon$ put the transition further from the worldline region.  Note that Eq. (\ref{eq:threshold}) is equivalent to the fractional residual form 
\begin{equation}
\frac{z_i-\tilde{z}(y_i)}{\tilde{z}(y_i)}\geq \epsilon-1\,.
\end{equation}

Linear fits to Eq. (\ref{eq:logpredict}) with fixed $\{\alpha,\lambda\}$ and varying $\ell$ do not conclusively confirm the theoretical predictions for the choices of $\epsilon$ attempted, which is not surprising given the discrepancies in Fig. \ref{fig:transition}.  Early results suggest
\begin{equation}
    x_* \sim B_1 \left(\frac{\tau \beta^{1/\alpha}}{\ell \eta^{1/\lambda}} \right)^{B_2}\,,
\end{equation}
where $B_1$ can vary by a few orders of magnitude and $B_2$ is roughly in the ballpark of 3.  However, for each case of $\{\alpha,\lambda\}$ we only have four or five different datasets, limiting the reliability of the fit results.  Furthermore, as shown in Fig. \ref{fig:transition}, for large values of $\ell$ there is very little data in the worldline regime to begin with, heavily skewing the estimates of $\ln x_*$.  The omission of the higher $l$ modes could exacerbate this problem.  One trial dataset using $l=0,1$ shows more data at smaller values of $x$, which could shift the location of the transition upon coarse-graining.  We leave for the future more careful analysis of the transition behavior.

\section{Outlook and discussion}
\label{sec:conclusion}
Large fluctuations of stress-tensor-like operators have a number of potentially observable and interesting effects, including fluctuating gravity waves~\cite{Wu:2011gk}, increased barrier penetration probabilities for charged particles~\cite{Huang:2016kmx}, alternative processes for false vacuum decay~\cite{Huang:2020bzb}, and greater variance of scattered photons in a low-temperature light scattering experiments~\cite{Wu:2020hrz}.  The extent to which these effects have physically observable manifestations depends on the likelihood of these large fluctuations, which can be investigated through the underlying probability distributions.  The asymptotic behavior of these distributions can be deduced from the high moments of the quadratic operators in question, although the operators need to be averaged in time alone or space and time for the moments to be finite.  Purely time-averaged operators have been discussed in two dimensions with Gaussian time sampling functions~\cite{Fewster:2010mc}, in four dimensions with Lorentzian time sampling functions~\cite{Fewster:2012ej}, and more recently in four dimensions with compactly supported time sampling functions~\cite{Fewster:2015hga}.  The latter two scenarios were numerically verified in Ref.~\cite{Schiappacasse:2017oqu}, providing confirmation of the high moments method.  However, a physical experiment takes place not only in a finite duration but also in a finite volume, motivating work on stress tensor operators averaged by compactly supported space and time sampling functions~\cite{PhysRevD.101.025006}.  In four dimensions, the probability distributions of spacetime-averaged operators asymptotically approach the worldline limit described by a purely time-averaged operator, $P(x)\sim e^{-x^{-\alpha/3}}$, before transitioning to a form that decays faster, $P(x)\sim e^{-x^\alpha}$, where $x$ is a dimensionless quantity proportional to the eigenvalues of the operator and the value of $\alpha$ determines the switch-on/off behavior of $f(t)$.

In this paper, we adapt the method developed in Ref.~\cite{Schiappacasse:2017oqu} for the case of spacetime-averaged operators.  The Minkowski vacuum state is generally not an eigenstate of an arbitrary normal-ordered quadratic operator averaged in space and time, so repeated measurements of the operator lead to different outcomes with different probabilities of occurence.  To construct the associated probability distribution, a Bogoliubov transformation is performed to find the eigenvalues and eigenkets.  The probabilities of measuring these eigenvalues in the vacuum state is then given by the squared inner products of the eigenkets with the vacuum state.  Choosing a suitable set of eigenkets, we can numerically construct the probability distributions by finding the eigenvalues and corresponding probabilities.

Numerically constructing the distribution for the spacetime average of $:\dot{\varphi}^2(t,\mathbf{r}):$, where $\varphi(t,\mathbf{r})$ is the massless scalar field, we find that similar outcomes can have wildly different probabilities of being measured.  As physical measurements are not precise enough to probe single eigenvalues, we argue that binning the data is a plausible resolution that produces smoother, more well-behaved data sets.  Alternative, rigorously developed coarse-graining methods may perhaps already exist in other fields, but in any case our procedure should capture the key qualities we expect from these averaged distributions.  Whether the bin sizes carry any physical meaning is speculated but better left for future investigation.

Our results show clear worldline and spacetime-averaged behavior with obvious transitions between the two limits, allowing analysis of the asymptotic behavior of the probability distributions.  Fitting to the asymptotic regions of the averaged probability distributions for the cases $\alpha=\lambda=0.5$; $\alpha=0.5$, $\lambda=0.7$; and $\alpha=\lambda=0.5$, we find that the decay rates in spacetime-averaged limits are consistent with prediction, whereas those of worldline limits are slightly lower than expected.  The latter inconsistency may result from limitations of the datasets and the approximations used to analyze the transition to and behavior in the spacetime-averaged limit.  When the full asymptotic form in the spacetime-averaged limit is predicted theoretically, we expect that a more careful analysis of the numerical data will find good consistency in both the worldline and spacetime-averaged limits.

In contrast, the predicted locations $x_*$ for the transitions between these two limits are somewhat inconsistent with data, though the qualitative behavior holds.  We do not have enough datasets to numerically analyze the transition behavior in detail, a problem compounded by the lack of data in the worldline region in some computations.  Preliminary analysis suggests the predicted power law behavior for $x_*$, Eq. (\ref{eq:xstar}), may hold, but we lack sufficient data to conclusively show this.  Though the effects of higher $l$ and $m$ modes are not well understood, we may expect nontrivial contributions because the spherical Bessel functions are nonzero as one moves away from the origin.  Such effects could shift the transition location, a speculation that appears plausible from some early trial runs.  Further exploration of the consequences of the $l$ and $m$ modes, perhaps in conjunction with more $n$ modes, may be worth pursuing in the future.

In addition to the transition behavior, another work in progress is the generalization to rectangular coordinates.  Although in this paper we work in spherical coordinates to take advantage of the spherical symmetry present in setup, spherical coordinate systems can be unwieldy in many contexts.  Computations in rectangular coordinates could allow for more straightforward applications in a variety of scenarios and perhaps even facilitate more sophisticated simulations, but the selection of modes can be tricky and more work remains to be done.  In spherical coordinates, a sufficient choice for mode selections is to set $l=m=0$, but the analogous choice in rectangular coordinates is not so clear.

\section{Acknowledgments}
We thank Chris Fewster for valuable discussions.   This work was supported by the Academy of Finland Grant 318319 and by the National Science Foundation under Grant No. PHY-1912545.\\
$^\dagger$\href{mailto:Peter.Wu610348@tufts.edu}{Peter.Wu610348@tufts.edu}\\
$^\ddagger$\href{mailto:ford@cosmos.phy.tufts.edu}{ford@cosmos.phy.tufts.edu}\\
$^*$\href{mailto:edschiap@uc.cl}{edschiap@uc.cl}\\

\bibliography{SpaceTimeAverage_2020} 
\end{document}